\documentclass[twocolumn,amssymb,superscriptaddress,floatfix,nobibnotes,aps,prx,superscriptaddress]{revtex4-2} 

\usepackage[T1]{fontenc}
\usepackage[utf8]{inputenc}
\usepackage{amsmath,amssymb}
\usepackage{newtxtext,newtxmath}
\usepackage[final]{microtype}
\usepackage{bm}
\usepackage{siunitx}
\usepackage{graphicx}
\usepackage{float}

\usepackage{booktabs}
\usepackage{enumitem}
\setlist{nosep,leftmargin=*}

\usepackage{caption}
\usepackage{subcaption}
\usepackage{hyperref}
\usepackage[nameinlink,noabbrev]{cleveref}
\crefname{equation}{Eq.}{Eqs.}
\Crefname{equation}{Equation}{Equations}
\crefname{figure}{Fig.}{Figs.}
\crefname{figure}{Figure}{Figures}
\crefname{table}{Table}{Tables}
\Crefname{table}{Table}{Tables}
\crefname{section}{Sec.}{Secs.}
\Crefname{section}{Section}{Sections}
\crefname{subsection}{Sec.}{Secs.}
\Crefname{subsection}{Section}{Sections}
\newcommand{\vect}[1]{\bm{#1}}
\newcommand{\dd}{\mathrm{d}}
\newcommand{\Ecoli}{\textit{E.\ coli}}

\graphicspath{{figures/pdf/}{figures/png/}{../figures/pdf/}{../figures/png/}{figures/}}

\begin{document}

\title{Embodied physical learning: a trained bacterial cell wall can guide organism behavior}

\author{Nada Elmeligy} \email{nnelmeli@syr.edu}
\affiliation{Physics Department, Syracuse University, Syracuse, NY 13244 USA}

\author{No Chen}
\affiliation{Physics Department, Syracuse University, Syracuse, NY 13244 USA}

\author{J. M. Schwarz} \email{jmschw02@syr.edu}

\affiliation{Physics Department, Syracuse University, Syracuse, NY 13244 USA}
\affiliation{Indian Creek Farm, Ithaca, NY 14850 USA}

\begin{abstract}
Physical learning studies how physical systems themselves can be trained to perform learning tasks using physical laws. For instance, disordered spring networks can be trained by local rules to route a mechanical signal from one site in the network to another. Disordered spring networks play key roles in many facets of cell biology, not just in terms of mechanics, but also in terms of motility. How then can training on the substrate itself impact the action of an organism, such as a bacterium? Here, we train a spring-network model of the bacterial cell wall and embed it in a motile agent. The stimulus is the load the cell generates by colliding with a surface. The trained wall routes that load across the body to distant flagellar motors, where it drives stator release, and the resulting turn biases the cell toward escape and delays subsequent surface contacts. On the first near-wall tumble after entrapment, the trained cell recovers $89\%$ of the advantage an ideal side-resolved controller provides. We call this arrangement embodied physical learning: the stimulus, the actuator, and the consequences all belong to the agent, and the training force scale is inherited from the agent's own sensing rather than imposed. An actuator with its own kinetics imposes a second, tighter constraint: the routed signal must switch the actuator within the stimulus window. We also find that behavior converges within a few hundred wall encounters, comparable to what a cell in sustained contact with structure experiences in a single generation, and that the routing survives continuous crosslink remodeling provided crucial, tension-bearing bonds are protected. Experimental tests of embodied physical learning are discussed. 
\end{abstract}

\maketitle
\raggedbottom
\setlength{\parskip}{0pt}

\section{Introduction}

Physical learning provides a precise language for asking whether matter can be trained to compute~\cite{SternMurugan2023,momeni2025training}. Contrastive learning approaches have helped lead the way in terms of algorithms~\cite{Scellier2017,scellier2021deep,Stern2021,Anisetti2023,li2024training,anisetti2023emergent, anisetti2024frequency, falk2025temporal}. Disordered electrical and elastic networks have served as the matter testbeds ~\cite{Dillavou2022,Altman2024}. As for the computational tasks, classification has been implemented in the disordered electrical networks~\cite{Dillavou2022}, and allosteric response in which there is a trained mechanical signal transmitted from one site (or edge) to another~\cite{Rocks2017, Pashine2019} has been implemented in disordered elastic networks~\cite{Altman2024}. However, these demonstrations were largely in abstract networks, where the input and output are imposed mechanical constraints, for example, and not signals an organism generates or uses. A recent cytoskeletal model moved closer to a biological setting: a coarse-grained cytoskeletal network with mechanosensitive proteins and active motors can implement contrastive learning, classify mechanical signals, and even train autonomously, without an external supervisor and under continuous turnover of its components~\cite{Banerjee2026}. Cells, therefore, contain molecular components that indeed support self-organized physical learning. 

In a swimming bacterium, navigation is traditionally assigned to intracellular chemotaxis signaling: a molecular cascade that acts as an internal memory, biasing motor switching rates and shaping run--tumble behavior~\cite{Macnab1972,Berg1972,Sourjik2002}. In this picture, the cell wall serves as passive mechanical infrastructure, providing structural integrity, but playing no active role in sensorimotor coordination. And yet the wall has the ingredients of a trainable network. It is disordered and filamentous~\cite{turner2018molecular}, its crosslink density is locally regulated by enzymatic remodeling~\cite{Vollmer2008}, and its bonds anchor the flagellar motors through a load-sensitive catch-bond interface~\cite{Nord2017}. Load applied at one motor is already, in principle, mechanical information transferred to another motor.  

What can such information be used for? Can it be used for coordination amongst motors to drive cell motility as a cell moves through a simple chemical environment (no chemical gradients) but a complex mechanical one? It has been noted that a trained cytoskeletal network has not yet been linked to an action performed by the cell~\cite{Banerjee2026}. Here, we make this link explicit by training a biological substrate through local learning rules, which may be more biologically plausible in various settings, to drive such an action. We call this link \emph{embodied physical learning}: a physical network whose input--output map is set by a local rule acting on its own degrees of freedom, deployed such that
\begin{enumerate}[label=(\roman*)]
\item the stimulus arises from an organism's (agent's) interaction with its environment;
\item the trained response drives an actuator with its own kinetics;
\item the actuated output alters the organism's (agent's) state and hence the future statistics of the stimulus.
\end{enumerate}

We call (i)--(iii) the deployment conditions. A further \emph{autonomy condition}, that training itself proceed without an external supervisor, is not yet implemented here. So for now, embodied physical learning here refers to embodiment of the learned physical response, not autonomous acquisition of that response. To date, trained elastic networks with imposed boundary
conditions~\cite{Rocks2017,Pashine2019,Stern2021} satisfy none of (i)--(iii): input, output, and scoring all sit outside any agent. The cytoskeletal model of Ref.~\cite{Banerjee2026} satisfies the autonomy condition, but its learned response remains internal to the network rather than driving an actuator that alters the cell's subsequent interactions with its environment. Thus, meeting all four conditions remains an outstanding goal. 

The deployment conditions concern functional requirements. However, embodied physical learning can also involve dynamical requirements. In general, several timescales may be relevant, including the number of training epochs required to acquire a prescribed response, the mechanical relaxation time of the network, the duration of the agent's interaction with its environment, and the actuator response time. If the environmental stimulus itself varies in time, its characteristic timescale introduces another consideration. Here, we explore the training scale, measured in epochs, as well as the interplay between actuator kinetics and the duration of the agent's interaction with the stimulus. We adopt an approximate hybrid model in which training and mechanical equilibration are treated quasistatically, while actuator kinetics and the agent's interaction with the environment are treated dynamically. We find that, for the bacterial wall-contact problem, condition (ii) introduces a timescale constraint: the transmitted mechanical signal must be sufficiently strong to drive the actuator to a functionally effective state before contact ends. Unlike an isolated trained network, whose performance is evaluated against a prescribed target amplitude, the embodied network must also produce an actuator response within the available interaction time. 

Given the three conditions for embodied physical learning, we train a two-dimensional, spring-network model of the peptidoglycan wall to route contact load across the cell body, then couple it to a swimming agent. On the first tumble after contact the trained cell escapes on $0.62$ of contacts against $0.43$ for an untrained wall, recovering $89\%$ of the advantage an ideal side-resolved controller provides. The advantage is robust in two ways we test directly, saturating early in training and surviving continuous crosslink turnover provided tension-bearing bonds are protected. In a central-force network, trainability itself dramatically increases right near the Maxwell isostatic point $z_c \approx 4$, so the same rigidity threshold that enables the unstressed wall to bear load also enables it to sustain a learned pathway. These properties allow the load-bearing wall to act as a mechanical steering pathway running parallel to chemotaxis, making the object of study a behavior the entire organism performs rather than an internal network response.

Our manuscript is organized as follows. In Sec. II, we introduce the elastic-network model of the peptidoglycan cell wall, the local stiffness-training algorithm, and its coupling to flagellar motor dynamics and bacterial motility. In Sec. III, we first demonstrate that training produces mechanical pathways that transmit contact forces across the cell wall, investigate how network connectivity and the rigidity transition constrain trainability, and show how these pathways bias bacterial motion away from confining surfaces. We then examine the timescale constraints imposed by actuator dynamics, the number of training cycles required to achieve functional behavior, and the robustness of learning under continuous crosslink turnover. In Sec. IV, we discuss the broader implications for embodied physical learning, the limitations of the model, and directions for future work.
\section{Model}
\begin{figure}[tbp]
  \includegraphics[width=\columnwidth]{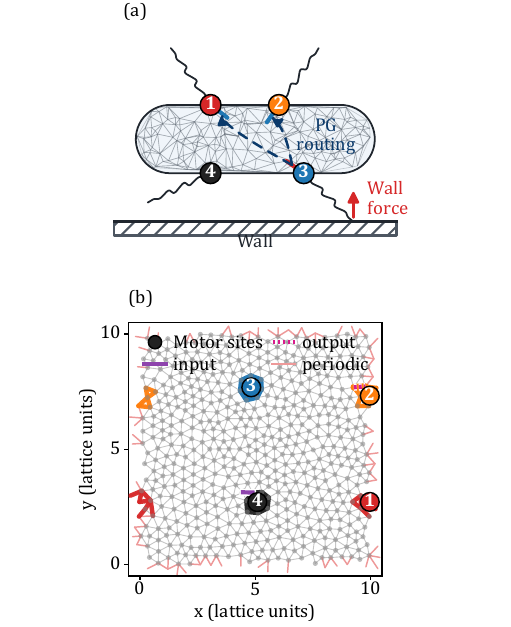}
\caption{\textbf{Model swimmer and peptidoglycan network.}
\textbf{(a)}~Wall contact loads motor-anchor edges on the impacted
side (red); the trained PG network routes the signal to far-side
readout edges (blue).
\textbf{(b)}~Periodic disordered network with 512 nodes, 1497 springs,
and mean coordination $z\approx5.85$, in a
$10\times10$ lattice-unit box. Colored circles identify the four
motors; anchor bonds follow their motor colors. Purple solid and
magenta dashed highlights mark designated input and output bonds,
respectively. Light-pink bonds cross the periodic boundary.
Stiffnesses are trainable; rest lengths remain fixed.}
  \label{fig:setup}
\end{figure}

\subsection{\Ecoli{} motility}\label{sec:motility}

The bacterial cell is a rod-shaped, \Ecoli{}-like swimmer; we model run and tumble by stochastic switching of each flagellar motor between CCW (run) and CW (tumble) at fixed rates calibrated to a $\sim 1\,$Hz tumble frequency~\cite{Berg1972} (calibration in S4.2, \cref{fig:calibration}, full parameter list in \cref{tab:bio-params}, S7). Since there are no chemical gradients in the medium, the gradient-sensing/adaptation pathway is not engaged (\cref{fig:setup}a). Flagellar motors are rotary machines whose filaments bundle to run the cell forward and leave the bundle during a tumble~\cite{Turner2000}. The flagellar motors are driven by stator units embedded in the inner membrane and anchored to the peptidoglycan (PG) cell wall through MotB~\cite{Santiveri2020}; in the model, each flagellar motor couples to the wall through a ring of edges where the stator units are anchored (the MotB--PG locus; motor placement detailed in S1).

\begin{figure}[tbp]
  \centering
  \includegraphics[width=\columnwidth]{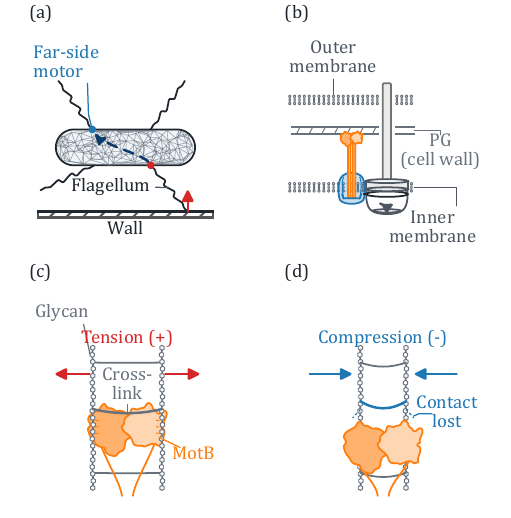}
  \caption{\textbf{Biological motivation and effective coupling.}
\textbf{(a)}~A flagellum contacts the wall; the dashed pathway
indicates the trained PG response.
\textbf{(b)}~Membrane-embedded stators anchor to PG through
MotB~\cite{Santiveri2020}.
\textbf{(c)}~Contact supplies a prescribed tensile input
$t_i^{\mathrm{in}}=cF_{\mathrm{wall},i}$ at one anchor edge.
The tensile sign and gain $c=1$ are model assumptions;
the input is independent of stator occupancy.
\textbf{(d)}~Far-side compression is hypothesized to disrupt
PG attachment and promote stator release.
Red denotes input; blue denotes output.
MotB-C profiles derive from the \textit{H.\ pylori} dimer,
PDB 3CYP~\cite{Roujeinikova2008}. Geometry is schematic,
not to scale.}
  \label{fig:mechanism}
\end{figure}

The mechanical stability of the flagellar motors depends on the continuous exchange and load-dependent retention of their stator units. The stators have a baseline rate of attachment and detachment to the wall, but load torque on the flagellar motor has been shown to activate a mechanosensitive pathway causing the stator units' detachment rate to fall off. This allows the stators to anchor for longer so the motors can push against the load~\cite{Lele2013, Tipping2013, Wadhwa2019}. Whereas that established pathway stabilizes stators locally at a loaded motor, here we ask whether contact load at one motor can propagate through the PG wall to destabilize stators at distant motors. This proposed allosteric pathway is summarized in \cref{fig:mechanism}.
When a rotating flagellum contacts the wall (a), the increased resistive load on the filament raises the reaction force transmitted through the membrane-embedded stators to their MotB--PG tether (b), pulling the local anchor crosslink taut. We therefore represent this local PG loading by a prescribed tensile input (c). The trained network produces compression at distant anchors (d), where we hypothesize enhanced stator detachment. The tensile input and compressive output therefore act at different sites.

Why is it advantageous to reduce stator occupancy at
motors on the side opposite the wall? In our
two-dimensional model, clockwise-rotating flagella
on the wall-facing side exert propulsive forces that
tend to reorient the bacterium away from the wall,
whereas those on the opposite side tend to reorient
it toward the wall. Suppressing the propulsive
contribution of the latter therefore favors escape.
We propose that the trained peptidoglycan network
accomplishes this by transmitting the mechanical load
generated at the impacted motor into compression at
stator anchors on the opposite side of the cell.
Through the hypothesized compression-dependent
detachment pathway, this compression reduces stator
occupancy and, consequently, the propulsive forces
of the far-side flagella, biasing reorientation
away from the wall.

\subsection{Actuator: Details of a proposed compression-release channel}\label{sec:actuator}

Our proposed compression-dependent stator detachment mechanism relies on the geometry of the interaction between MotB and the peptidoglycan network. The MotB stator domain anchors the motor to the peptidoglycan through a C-terminal OmpA-like domain~\cite{Santiveri2020, Deme2020}. Rather than binding a single chain, the stator dimer is modeled to bridge two adjacent glycan strands, with a binding surface whose span closely matches that of an extended peptide crosslink~\cite{Roujeinikova2008} (\cref{fig:mechanism}b). Inserting into this mesh requires a conformational rearrangement to expose the binding residues~\cite{Reboul2011,Kojima2018}. Direct binding of this activated domain to purified PG has been demonstrated~\cite{Kojima2018}; because this OmpA-like fold and PG-binding motif are highly conserved, we assume this geometry to hold for \Ecoli{}.

We, therefore, treat the anchoring process as fundamentally geometric: it requires the local crosslink to present a span wide enough for the domain to bridge. That the detachment rate is set at this interface rather than at the rotor is consistent with its independence of rotation direction: clockwise and counterclockwise motors have different torque--speed curves, but identical off-rates at matched torque~\cite{Wadhwa2021}. Increasing load on the motor pulls this crosslink taut, which in our geometric reading corresponds to the established load-dependent catch bond that stabilizes the stator~\cite{Nord2017,Chawla2017}. The complementary regime is, to our knowledge, untested: if the crosslink is instead compressed below that span, the geometric match is broken and the stator should fail to hold.

We adopt this as a hypothesis about the actuator response (\cref{fig:mechanism}d): where the trained network drives the local anchor crosslink into compression, the stator detachment rate rises. This assumes the dimensional match is a tight constraint. The actuator is the stator occupancy at each motor. We describe motor $i$ by a continuous mean occupancy $S_i\in[0,N_{\text{stat}}]$, where $N_{\text{stat}}$ is the maximum number of bound stators per motor. Its dynamics obey
\begin{equation}
  \frac{\dd S_i}{\dd t} = k_{\text{on}}(N_{\text{stat}} - S_i) - k_{\text{off},i}\,S_i,
  \label{seq:stator}
\end{equation}
with on-rate constant $k_{\text{on}}$ so all directional asymmetry enters through the off-rate $k_{\text{off}}$. The off-rate is the sum of two independent detachment channels,
\begin{equation}
  k_{\text{off},i}
  = k_{\text{off}}^{\text{LS}}\,\exp\!\bigl(-\beta\,F_{\text{stat},i}\,\delta_{\text{LS}}\bigr)
  + k_{\text{off}}^{\text{geo}}\,\Bigl[\exp\!\bigl(\beta\,\delta_{\text{geo}}\,\sigma_{\text{comp},i}\bigr) - 1\Bigr],
  \label{seq:koff}
\end{equation}
where $\sigma_{\text{PG},i}$ is the signed PG readout-edge tension at motor $i$, positive in tension and negative in compression, and $\sigma_{\text{comp},i}=\max(0,-\sigma_{\text{PG},i})$ is its compressive magnitude. The per-stator anchor force $F_{\text{stat},i}$ is the baseline reaction force from driving the filament against viscous fluid drag ($\sim 9\,$pN per stator in open field, determined by the motor's torque--speed operating point in S4.4 and uncoupled from wall collision), whereas $\sigma_{\text{comp},i}$ is supplied by the PG network during wall contact; both have force units (pN). Here $\beta=(k_BT)^{-1}$, $k_{\text{off}}^{\text{LS}}$ is the first channel's off-rate at zero anchor force, and $k_{\text{off}}^{\text{geo}}$ is the rate scale of the geometric channel. The effective Bell distances $\delta_{\text{LS}}$ and $\delta_{\text{geo}}$ have units of length (nm), rather than representing the instantaneous crosslink lengths; the former is taken from prior measurements and the latter is a model parameter. The first term describes load-stabilized stator retention, with increasing anchor force lowering the off-rate. The second describes hypothesized compression-induced release and vanishes when $\sigma_{\text{comp},i}=0$, leaving the baseline occupancy dynamics unchanged in the absence of compression. Because $\sigma_{\text{comp},i}$ is rectified at zero, this channel is one-directional. Parameter values and their justifications are given in \cref{tab:bio-params}, with the basis for each channel detailed in S4.3.

\subsection{PG Layer as a spring network}

The PG layer making up the cell wall is represented as a two-dimensional disordered spring network. To ensure that there are no underlying translational symmetries in the architecture of the disordered network, the architecture is based on a bidisperse jammed packing configuration in which particle contacts are springs. Moreover, we impose periodic boundaries in both directions (a torus; full construction in S1, network shown in \cref{fig:setup}b). The disorder reflects the disordered, filamentous architecture of PG~\cite{turner2018molecular}, and the packing gives $z\approx5.85$, above the isostatic point $z_c=4$, so the unstressed network is rigid (\cref{fig:sweep}).

The peptidoglycan layer behaves elastically on timescales shorter than cell growth~\cite{Yao1999,Thwaites1989,Rojas2018,Deng2011}, thereby motivating its representation as an elastic spring network~\cite{Nguyen2015,Gumbart2014,Vollmer2010}. Nodes $\vect{r}_i$ are the network sites; edges $\langle ij \rangle$ connect nearby nodes. Each edge carries a stiffness $k_{ij}$ (the trainable degree of freedom) and a rest length $\ell^0_{ij}$, which is held fixed during training. Biologically speaking, the trainable stiffness corresponds to enzymatic remodeling of
cross-link density, built by transpeptidases and cleaved by
endopeptidases~\cite{Singh2012,Typas2010, Wheeler2015} under spatial regulation by the Lpo adaptors~\cite{Typas2010}.

Mechanical equilibrium in the network is obtained via the quasistatic approximation (S1) by minimizing the total stretching energy
\begin{equation}
  E = \sum_{\langle ij \rangle} \tfrac{1}{2}\, k_{ij}\bigl(|\vect{r}_i - \vect{r}_j| - \ell^0_{ij}\bigr)^2,
  \label{eq:energy}
\end{equation}
subject to boundary conditions and any applied edge forces.

\subsection{Stiffness training by a local feedforward-feedback rule}

We train the network to transmit mechanical stress from an
input edge to a designated output edge by adjusting the
stiffnesses of the trainable springs. The input and output edges are held fixed and a local learning rule, derived from Multi-Mechanism Learning~\cite{Anisetti2023,anisetti2024frequency} is used for training. Each training cycle
involves two mechanical equilibrations, corresponding to
feedforward and feedback states.

In the Feedforward (IN) state, a tension corresponding to the mechanical load of a flagellar-wall collision is applied
to the input edge, and the network is mechanically
equilibrated. The resulting signed tension
$t_b^{\rm FF}$ is recorded for every bond $b$.
In particular, the tension $t_e^{\rm F}$ at each target
output edge $e$ is compared with its desired value
$t_e^D$. This difference determines the loading
applied in the feedback state,
\begin{equation}
t_e^{\rm FB} = t_e^D - t_e^{\rm F},
\qquad e\in\mathcal{T},
\label{eq:feedback_loading}
\end{equation}
where $\mathcal{T}$ denotes the set of target edges.
In the feedback state, the input loading is removed such that free tensions $t_e^{\rm F}$ are used, 
the prescribed feedback tensions are applied at the
target edges, and the network is again mechanically
equilibrated. The resulting signed tension
$t_b^{\rm FB}$ is recorded for every trainable bond.

The stiffness of each trainable bond is updated
according to the product of its feedforward and
feedback tensions,
\begin{equation}
\Delta k_b =
-\gamma_l\,t_b^{\rm FF}t_b^{\rm FB},
\label{eq:stiffness_update}
\end{equation}
where $\gamma_l$ is the learning rate. Bonds experiencing
opposite-sign tensions in the two states are stiffened,
whereas bonds experiencing same-sign tensions are
softened. 

For the present task, which requires a
compressive response to a tensile input, this rule
promotes the formation of stiffened pathways that
transmit the desired mechanical response. The bond tensions are signed scalars, positive in
tension and negative in compression. After each update,
the stiffnesses are constrained to the interval
$[k_{\min},k_{\max}]$, while the bond rest lengths
remain fixed throughout training. Further details
of the training protocol are provided in Sec.~S2
of the SI.

The local feedforward-feedback training rule suggests a possible mechanism by which \Ecoli{} could adapt its peptidoglycan network to improve its response to environmental stimuli. While mechanical forces arising from flagellar contact with a boundary generate the feedforward field, forces transmitted from the flagellar motors back into the peptidoglycan network could provide a feedback field. Local remodeling of the network in response to these coupled mechanical stresses could then reinforce pathways that coordinate environmental sensing with motor activity. Although the biological implementation of such a learning rule remains to be established, our results suggest a route by which the bacterium could train its own cell wall through mechanically mediated feedback.

To bias motility, the network must be trained to transmit mechanical information between all motor sites. The goal is routing mechanical stress from an impacted flagellar motor to the anchor edges of the motors on the free side of the cell. Our \Ecoli{} model has four flagellar motors, so we cyclically train eight distinct input--output tasks. In each task, one motor acts as the ``impacted'' input and a single motor on the opposite side of the cell acts as the ``non-impacted'' target; the eight tasks enumerate every pairing of the four motors with the two motors on their opposite side.

To facilitate the escape behavior introduced in \cref{sec:motility}, the training objective is an opposite-sign allosteric response. We assume that wall contact produces tensile loading at the impacted flagellar motor's peptidoglycan anchor (\cref{fig:mechanism}c), matching the sign of the taut anchor crosslink in our geometric reading of load stabilization. The network must therefore be trained to transmit this local tensile perturbation ($+0.30\,$pN at the input edge) into compression at motor anchors on the opposite side of the cell (a targeted compressive response of $-0.12\,$pN at the output edge). As detailed in \cref{sec:actuator}, this routed compression drives far-side stator detachment, which the near-wall assay (\cref{sec:assay}) utilizes to bias the cell's reorientation away from the surface.

The training input is set by the maximum contact force generated by the swimmer itself, so the $+0.30\,$pN magnitude is not a free parameter. As detailed in the near-wall assay (\cref{sec:assay}), this value is the kinematic ceiling of the contact signal a flagellum generates against a surface, which coincides with the propulsive thrust of a single filament (Sec.~S4.5 of the SI). Thus, the network is trained at the largest signal its own sensor can deliver rather than at an imposed scale. Given that the network operates in the linear elastic regime across this physiological force range (Sec.~S6 of the SI), training at this kinematic ceiling ensures that the targeted transfer ratio is strictly preserved for all intermediate contact loads the cell encounters.

We train each input–output pairing as a separate task to avoid degenerate solutions. Each task carries a single target edge. An earlier protocol used four tasks, each designating both opposite-side motors as targets within one cycle. The global error fell, but the resulting solution was degenerate: the network delivered a ratio near $-0.9$ to one of the two targets and essentially nothing to the other, satisfying the aggregate objective without routing to both. Scoring each pairing on its own cycle removes that degeneracy, at the cost of requiring one stiffness field to satisfy eight tasks rather than four.

 The target compression is constrained by the need to satisfy all eight training tasks. Accordingly, we set the target magnitude to \(-0.12\) pN, corresponding to a tension transmission ratio of \(-0.4\), so that both tasks in which a given motor acts as input can converge together. This per-task ratio of $-0.4$ is measured at one designated output edge with a single input edge driven. It is distinct from the transfer ratio $r = 0.905$ reported for the near-wall assay (Results), which measures the mean compressive readout at a far-side motor per unit contact force when both motors on the impacted side are loaded simultaneously, so their linear elastic contributions superpose.

\subsection{Near-wall assay}\label{sec:assay}

\begin{figure}[tbp]
 \centering
 \includegraphics[width=\columnwidth]{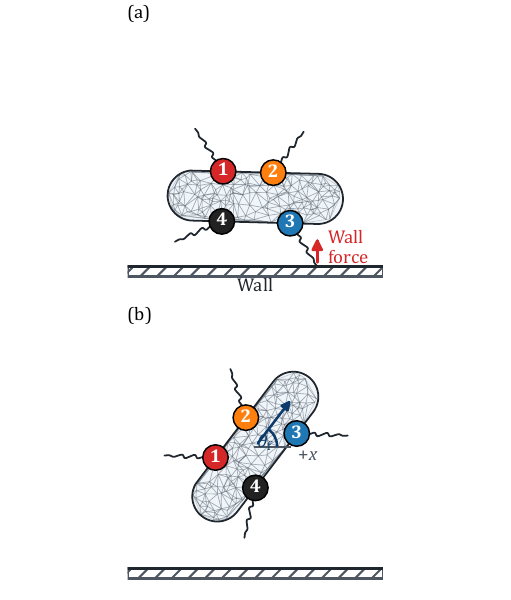}
\caption{\textbf{From wall contact to directed reorientation.}
\textbf{(a)}~Flagellum--wall contact loads a designated input
edge at the impacted motor (red). The trained network produces
compressive readouts at far-side motor anchors.
\textbf{(b)}~Compression increases far-side stator detachment,
creating an occupancy asymmetry $\Delta S$ that biases the
summed flagellar force. The cell turns by $\Delta\theta=\gamma$,
giving the post-tumble heading
$\theta_{\mathrm f}=\theta_{\mathrm i}+\gamma$.
The illustrated turn meets the trap-release criterion.
Without contact, the PG readout resets to zero.
Population statistics are shown in \cref{fig:pop-escape}.}
  \label{fig:run}
\end{figure}

We investigate the role of the cell wall stress propagation in mechanosensitivity by looking at the bacterial cell interaction with a wall. \Ecoli{} swimming near a wall becomes stably trapped in the near-surface region through a combination of hydrodynamic and steric interactions~\cite{Bianchi2017}. Whether tumbling provides an effective escape route is unsettled: surfaces have been reported to suppress tumbling and confine reorientations parallel to the wall, making it ineffective~\cite{Molaei2014}, whereas other work finds tumbling to be the dominant escape mechanism~\cite{Junot2022}. Here, we propose that the bacterium can use {\it mechanical information} generated by wall contact to coordinate flagellar motors on opposite sides of the cell, the impacted and non-impacted sides, to bias its subsequent motion away from the wall.

We model the bacterium as a two-dimensional swimmer in a
channel bounded by two parallel walls. Near-surface retention
is implemented through a trap-and-release rule adapted from
Ref.~\cite{Junot2022}, rather than by explicitly resolving
hydrodynamic interactions with the wall. We model any wall contact as a simple Hookean spring load
$F_{\text{wall},i}\ge0$ at motor $i$ (S5.1). It is transferred to a single designated input edge within the ring of anchor edges surrounding each motor (S1) as the prescribed tension $t_i^{\text{in}}=cF_{\text{wall},i}$, with $c=1$ (\cref{fig:mechanism}c). The peptidoglycan network transmits the resulting stress to flagellar motor anchors. 
In the trained network, compression at motor anchors on the opposite side of the cell increases stator detachment through
\cref{seq:koff} (\cref{fig:mechanism}d), producing an asymmetry in stator occupancy. The
rotation speed of motor $i$ depends on its stator occupancy
$S_i$ according to
\begin{equation}
\omega_i =
\frac{S_i \tau_{\mathrm{stat}}}
{\zeta_r + S_i \tau_{\mathrm{stat}}/\omega_{\max}},
\label{eq:motor_speed}
\end{equation}
where $\tau_{\mathrm{stat}}$ is the stall torque per stator,
$\zeta_r$ is the rotational drag coefficient, and
$\omega_{\max}$ is the zero-load motor speed. Motor switching between counterclockwise (CCW) and clockwise
(CW) rotation is governed by the stochastic switching model
described in Sec.~S4 of the SI. A tumble is initiated when at
least two of the four flagellar motors rotate clockwise.
The stator occupancy influences the motor dynamics through
Eq.~\eqref{eq:motor_speed}, while the switching model
determines the motor states and the occurrence of tumbles (\cref{fig:run}a).

To determine the bacterium's post-tumble direction, we follow
the simplified force-summation approach of Dvoriashyna and
Lauga~\cite{Dvoriashyna2021}, adapted to two dimensions. The bundled,
counterclockwise-rotating flagella contribute a propulsive
force along the cell axis, while clockwise-rotating flagella
contribute forces determined by their orientations. In the cell
frame, with $\hat{\mathbf e}_x$ along the pre-tumble cell axis, the
total propulsive force and the resulting signed reorientation angle are
\begin{equation}
\mathbf{F}_{\mathrm{tot}}
=
f_{\mathrm{bun}}\hat{\mathbf e}_x
+
\sum_{i\in\mathrm{CW}} f_i\hat{\mathbf n}_i,
\qquad
\gamma =
\operatorname{atan2}
(F_{\mathrm{tot},y},F_{\mathrm{tot},x}),
\label{eq:post_tumble_heading}
\end{equation}
where $f_{\mathrm{bun}}$ is the bundled flagellar
propulsive force, $f_i$ is the propulsive force of clockwise-rotating
flagellum $i$, and $\hat{\mathbf n}_i$ specifies its direction. At the end of a tumble, the cell reorients by
$\Delta\theta=\gamma$, so that its post-tumble
heading relative to the wall is
$\theta_{\rm f}=\theta_{\rm i}+\gamma$.
This heading determines whether the cell escapes
the near-wall region, as described below.
Because each flagellum's propulsive force is proportional to its motor speed ($f_i \propto \omega_i$, Sec.~S4.5 of the SI) and is restricted to point into the body half-plane~\cite{Dvoriashyna2021}, a clockwise flagellum on the wall-facing side pushes the cell body away from the wall, whereas one on the far side pushes it toward the wall. When the trained network compresses the far-side anchors and strips their stators ($S_i \to 0$, and thus $\omega_i, f_i \to 0$), the wall-ward push from far-side clockwise flagella is suppressed, leaving the away-from-wall push from the impacted side to bias the reorientation.

Finally, a bacterium is considered to escape the near-wall
region if its post-tumble heading allows it to travel beyond
the trapping distance $d_{\mathrm{trap}}$ within one mean run
time $\tau_{\mathrm{run}}$. For a heading now making an angle
$\theta_{\rm f}$ with the wall, escape requires
\begin{equation}
v_{\mathrm{run}}\tau_{\mathrm{run}}\sin\theta_{\rm f}
\geq d_{\mathrm{trap}},
\qquad
\sin\theta_c
\equiv
\frac{d_{\mathrm{trap}}}
{v_{\mathrm{run}}\tau_{\mathrm{run}}},
\label{eq:escape_criterion}
\end{equation}
where $v_{\mathrm{run}}$ is the swimming speed. The simulation
releases the bacterium when its heading points away from the
wall and $|\sin\theta_{\rm f}|>\sin\theta_c$. Otherwise, it remains
trapped until a subsequent tumble. Further details of the
motor dynamics, flagellar forces, and near-wall interactions
are provided in the SI.

\begin{figure*}[tbp]
 \centering
  \includegraphics[width=\linewidth]{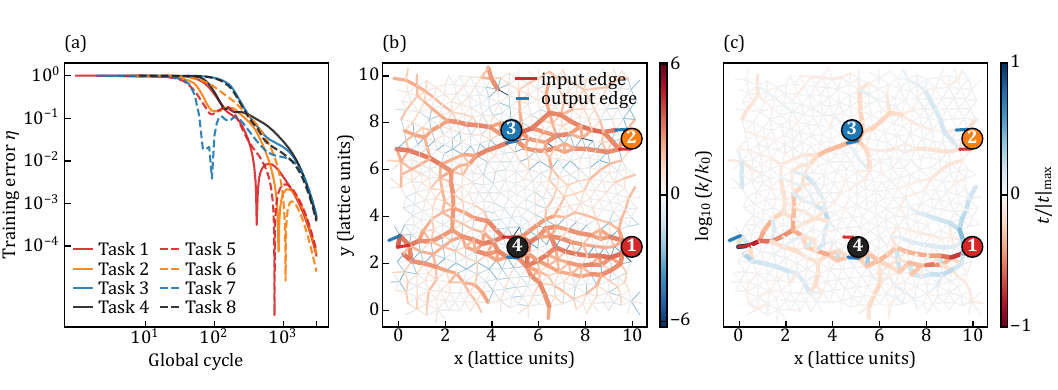}
  \caption{\textbf{Training produces structured mechanical pathways.}
\textbf{(a)}~Normalized training error $\eta$ versus global cycle
during cyclic presentation of eight input--output motor tasks.
\textbf{(b)}~Final stiffness map $\log_{10}(k/k_0)$, showing
heterogeneous remodeling from the initially uniform network.
\textbf{(c)}~Normalized feedforward bond tension
$t/|t|_{\max}$ for one input load.
Red and blue highlights identify the input and target readout
edges, respectively. The trained stiffness field organizes the
mechanical response into pathways connecting the loaded motor
to a far-side anchor. All panels describe the same trained network.}
  \label{fig:training}
\end{figure*}

\section{Results}
We now examine how the three conditions for embodied physical learning are satisfied in the trained bacterial system.


\subsection{Training mechanical pathways: The role of network rigidity}

We first investigate whether training can establish the prescribed mechanical pathways in the peptidoglycan network and how network rigidity constrains their formation. Before training, all bonds in the peptidoglycan network share an identical uniform stiffness $k_0$ and so the stress spreads more uniformly throughout the network. After cyclic training across the eight impact tasks, this uniform transmission breaks down and the training error $\eta$ (defined in S2) drops steadily with the global cycle (\cref{fig:training}a). As shown in \cref{fig:training}(b), the functional tuning is large and spatially heterogeneous: training drives nearly all bonds away from their initial stiffness $k_0$, with the largest changes spanning several orders of magnitude. These stiffened bonds form connected force chains linking the input and target regions, physically locking in the transmission pathways required to route a localized impact across the cell body. The functional consequence appears in the following sections: this remodeling raises the transfer ratio from a contact-side load to the far-side readout from $0.0056$ to $0.905$. Of course, this remodeling demands large changes in the spring stiffnesses, which will be addressed in the Discussion. 

\begin{figure}[tbp]
 \centering
  \includegraphics[width=\columnwidth]{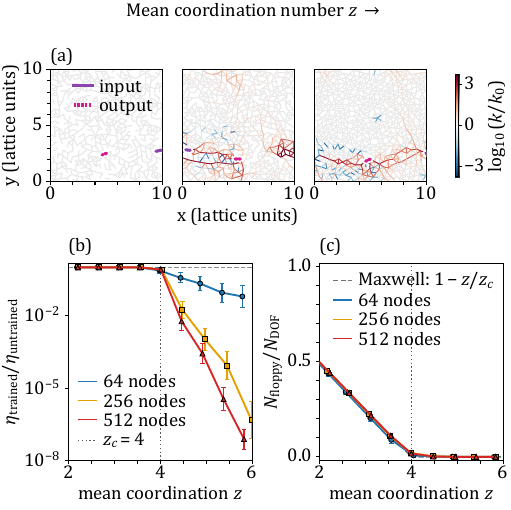}
 \caption{\textbf{Trainability near the rigidity threshold.}
\textbf{(a)}~Trained stiffness maps $\log_{10}(k/k_0)$ at three
mean coordinations $z$. Purple solid and magenta dashed highlights
identify input and target readout bonds, respectively.
\textbf{(b)}~Trained-to-untrained error ratio
$\eta_{\mathrm{tr}}/\eta_{\mathrm{un}}$ versus $z$ for
$N=64$, 256, and 512 nodes.
\textbf{(c)}~Floppy-mode fraction
$N_{\mathrm{fl}}/N_{\mathrm{DOF}}$, compared with the Maxwell
counting estimate $1-z/z_c$, with $z_c=4$.
The improvement in training performance occurs near this
central-force rigidity estimate.}
  \label{fig:sweep}
\end{figure}

The formation of these trained pathways is,
however, constrained by network rigidity. Inspired by the natural variation in bacterial cell wall topology and crosslinking density~\cite{Turner2013,Vollmer2008,Huang2008}, we ask how physical trainability depends on network connectivity. To test this, we systematically vary the bond occupation probability $p$ to set the mean coordination $z$ (protocol in S3). \Cref{fig:sweep} (top row) shows trained stiffness maps at three values of $z$. Below the critical coordination $z_c = 4$, no stiffened pathway forms; above $z_c$, a structured transmission channel appears. Quantitatively (\cref{fig:sweep}, bottom-left), the ratio of trained to untrained task error $\eta_{\text{tr}}/\eta_{\text{un}}$ stays near $1.0$ below $z_c$, meaning training fails entirely, and drops sharply above it. This drop steepens as the system size increases ($N = 64, 256, 512$), pointing to a critical transition.

This onset of trainability is consistent with the onset of rigidity approximated by the Maxwell prediction. As shown in \cref{fig:sweep} (bottom-right), the floppy-mode fraction $N_{\text{fl}}/N_{\text{DOF}}$ follows the Maxwell prediction $1 - z/z_c$ and vanishes right near $z_c = 4$. Below $z_c$, floppy modes prevent a trained pathway from holding tension; above $z_c$, the rigid network provides the mechanical support necessary to lock in the pathway. The ability to tune responses in disordered spring networks, and the requirement that connectivity exceeds the isostatic point, aligns with prior studies on designed allostery and multifunctional materials~\cite{Goodrich2015,Rocks2019}. 

Rigidity thus constrains both the structural and information-processing roles of the cell wall. Below $z_c$, the central-force network cannot reliably establish the desired mechanical pathways through training, whereas above $z_c$, rigidity provides the mechanical support necessary for their formation. The near coincidence of the rigidity and trainability thresholds suggests that the same network architecture that provides structural integrity can also support trainable mechanical responses.

\subsection{The trained wall steers the swimmer out of the trap}

\begin{figure}[tbp]
 \centering
  \includegraphics[width=\columnwidth]{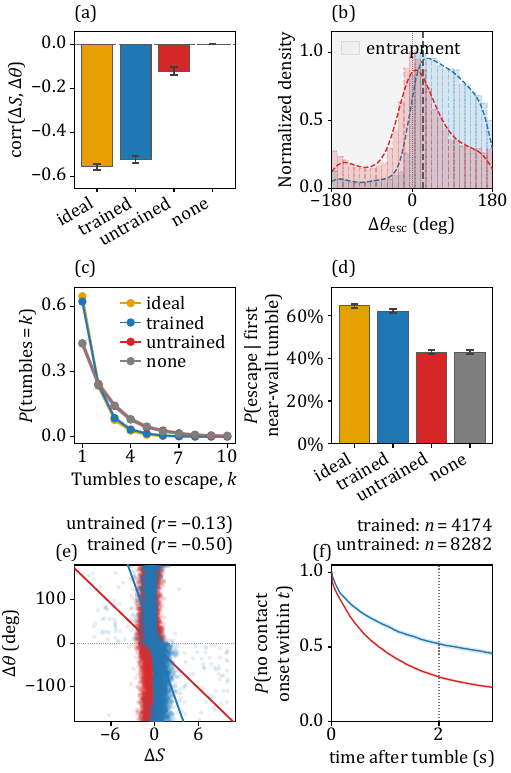}
\caption{\textbf{Population escape across four wall-response variants.}
Results from 100 independent trials per variant.
\textbf{(a)}~Trial-mean correlation between stator asymmetry $\Delta S$
and tumble angle $\Delta\theta$ (95\% CI).
\textbf{(b)}~Escape-aligned heading changes; shading marks the
entrapment zone ($|\Delta\theta|<24^\circ$).
\textbf{(c)}~Distribution of tumbles required to escape.
\textbf{(d)}~Escape probability at the first near-wall tumble.
Error bars in (c,d) are Wilson 95\% CIs.
\textbf{(e)}~Pooled $\Delta S$ versus $\Delta\theta$ for trained
(blue) and untrained (red) networks.
\textbf{(f)}~Probability of no new contact onset within time $t$
after a non-escaping tumble.}
  \label{fig:pop-escape}
\end{figure}

To test whether these trained mechanical pathways can guide navigation, we embed the trained network as the cell wall of a swimmer confined in a horizontal channel with top and bottom walls. We then route the signal from flagellar motor contact with the wall to the designated stator attachment edge on the trained network. Upon wall contact, the impacted side motors are loaded, and their anchor edges carry the input signal (\cref{fig:run}a). The trained network transmits this signal across the body to the motors on the far side where trained output signal shifts the local anchor tension, compressing the local PG edge, which increases the stator detachment rate. The resulting asymmetry in propulsive forces biases the post-tumble heading away from the contacting wall (\cref{fig:run}b).
The population statistics testing this causal chain are shown in \cref{fig:pop-escape}.

To isolate the effect of training, we compare the
trained wall with three reference cases in an
otherwise identical near-wall assay (S5.3).
First, the {\it untrained network} has the same topology,
geometry, and motor--edge assignments as the
trained network, but retains its initial uniform
stiffnesses. Second, the {\it no-network control} removes
mechanical signaling between motors by setting
the readout tensions to zero. Finally, the {\it ideal
controller} provides an upper benchmark for
side-resolved steering: it detects which side of
the cell contacts the wall and directly regulates
the motors, without losses from mechanical
transmission. Unlike the trained wall, which can
only increase stator detachment through compression,
the ideal controller can both stabilize stators
on the loaded side and destabilize those on the
opposite side (Model).

Across the population, the steering is sign-based: because far-side stator detachment saturates rapidly upon contact (as shown in the next section), the loaded side sets the direction of the escape turn rather than its magnitude. The trained network reproduces the ideal controller's sign-based coupling, whereas the untrained and no-network controls do not (\cref{fig:pop-escape}a,e; numerical values and confidence intervals are given in the figure caption). To combine left- and right-impact events onto a single axis, we use the escape-aligned heading change $\Delta\theta_{\mathrm{esc}} = \Delta\theta \cdot \mathrm{sign}(\theta_{\mathrm{req}})$; a positive value is a turn toward open space (\cref{fig:panel-d-delta}). On the first near-wall tumble after entrapment the trained network escapes on $0.62$ of events, against $0.43$ for the baselines and $0.65$ for the ideal controller, recovering $89\%$ of the ideal advantage (\cref{fig:pop-escape}d). The trained wall also delays the next contact even when escape fails, closing condition (iii) (\cref{fig:pop-escape}f). Pooled escape fractions, pooled coupling values, sample sizes, and the carry-over analysis are given in the figure caption and S5.4.

The two baselines are behaviorally indistinguishable because the untrained lattice transmits more than two orders of magnitude less than the trained wall (transfer ratio $0.0056$ against $0.905$), leaving a far-side readout of ${\sim}10^{-3}\,$pN and a stator asymmetry of $|\Delta S|\lesssim0.4$ against ${\sim}10$ for the trained wall (\cref{fig:steer-saturation}a). The untrained control establishes that training matters, while the no-network control establishes that the untrained mechanical coupling is too weak to produce an appreciable behavioral effect. The remaining difference between the trained
wall and the ideal controller arises primarily
from the distribution of post-tumble angles
rather than limitations in mechanical
transmission (\cref{fig:tumble-decomp}).

\subsection{The trained response must act within the stimulus window}
\begin{figure}[tbp]
 \centering
  \includegraphics[width=\columnwidth]{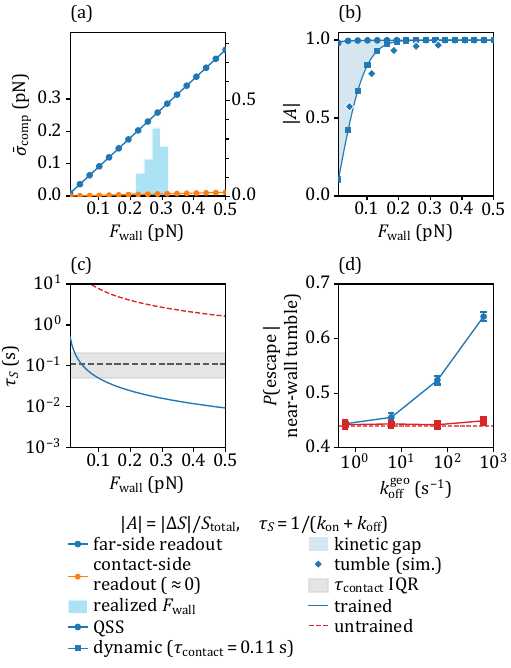}
  \caption{\textbf{Elastic routing and the kinetic operating window.}
Trained (blue) and untrained (red) networks.
\textbf{(a)}~Far-side compressive readout versus contact-side
load $F_{\mathrm{wall}}$; histogram shows realized per-motor
loads at tumble.
\textbf{(b)}~Normalized stator asymmetry $|A|$: steady state
(circles), response within a $0.11\,$s contact (squares),
and realized tumble medians (diamonds).
\textbf{(c)}~Stator relaxation time
$\tau_S=(k_{\mathrm{on}}+k_{\mathrm{off}})^{-1}$;
dashed line marks the median contact duration ($0.11\,$s).
\textbf{(d)}~Escape probability per near-wall tumble versus
geometric detachment-rate scale $k_{\mathrm{off}}^{\mathrm{geo}}$.}
  \label{fig:steer-saturation}
\end{figure}

Training enhances the transmission of mechanical signals through the cell wall without increasing the force available to drive them. Specifically, training raises the ratio of far-side compressive readout to contact-side load, $r \equiv \sigma_{\text{comp}}/F_{\text{wall}}$, from $0.0056$ to $0.905$, a $162$-fold increase, turning the network into a directed elastic router (\cref{fig:steer-saturation}a). The load itself is capped: a flagellum in Stokes flow cannot press on a surface harder than it can propel the cell, so $F_{\text{wall}}$ saturates near the thrust of a single filament (S4.5, S5.1). At this maximum load, the cell can increase the routed compression $\sigma_{\text{comp}} = r\,F_{\text{wall}}$ only by increasing $r$. Therefore, training changes not the force available, but the fraction of it that arrives at the far-side anchor.

Whether the routed compression can steer the cell depends on {\it both} the magnitude and the timescale of the stator response. Since \cref{seq:stator} is a first-order relaxation, it has both an endpoint and a characteristic time to reach it, and the off-rate sets both. Over the range of contact loads considered here ($\beta\delta_{\text{geo}}\sigma_{\text{comp}} \ll 1$), the geometric channel in \cref{seq:koff} is effectively linear in the routed compression, $k_{\text{off}}^{\text{geo}}[\exp(\beta\delta_{\text{geo}}\sigma_{\text{comp}})-1] \approx \kappa\,\sigma_{\text{comp}}$ with slope $\kappa \equiv k_{\text{off}}^{\text{geo}}\beta\delta_{\text{geo}} = 216\,\text{s}^{-1}\text{pN}^{-1}$. Thus, the compression-induced increase in the off-rate is $\kappa rF_{\text{wall}}$, coupling the mechanical transmission efficiency $r$ to the stator response (\cref{fig:steer-saturation}).

Reaching a large steady-state asymmetry is necessary for steering, but the actuator must also switch before the collision ends. The steady-state endpoint is the normalized asymmetry $A=(S_L-S_R)/(S_L+S_R)$. With $S \propto k_{\text{on}}/(k_{\text{on}}+k_{\text{off}})$ on each side, $|A|\ge0.99$ requires $k_{\text{off}}\gtrsim2\,\text{s}^{-1}$, which the trained pathway supplies just above the contact gate (\cref{fig:steer-saturation}b). The time is set by the relaxation rate $k_{\text{on}}+k_{\text{off}}$, and the asymmetry is expressed within a contact only when
\begin{equation}
  \tau_S = \frac{1}{k_{\text{on}}+k_{\text{off}}} \lesssim \tau_{\text{con}},
  \label{eq:switch}
\end{equation}
with $\tau_{\text{con}}$ the time a flagellum spends above the contact gate (median $0.11\,$s; S5.1). This kinetic requirement is tighter, $k_{\text{off}}\gtrsim9\,\text{s}^{-1}$, so the trained network must satisfy the contact window rather than only the steady-state endpoint. The trained wall determines whether the off-rate falls inside this window.

The distinction between steady-state and finite-time responses explains why the trained
wall steers the swimmer whereas the untrained wall does not. At the operating load, the trained pathway satisfies this window and the untrained
pathway does not (\cref{fig:steer-saturation}b,c) and so the 
two conditions are not redundant. Given unlimited time, the untrained wall would
reach $|A|=0.95$, ample asymmetry to steer with; within a contact it reaches
$0.02$. Its $162$-fold deficit in transmission matters only because the stimulus
ends. Nor can it compensate by pressing harder, given the maximum load (\cref{fig:steer-saturation}a). Training does not alter the intrinsic stator kinetics; rather, it increases the transmitted compression sufficiently for those kinetics to produce a substantial response during a collision.

The kinetic constraint also determines whether the trained mechanical pathway can improve the probability of escape. Escape follows this condition directly, and because the geometric off-rate is a modeling choice rather than a measured quantity (S4.3) we vary it. Holding the contact load at the physiological scale, steering and escape turn on between $k_{\text{off}}^{\text{geo}}=60$ and $600\,\text{s}^{-1}$: at $600\,\text{s}^{-1}$ the trained network reaches $P(\mathrm{escape}\mid\mathrm{near\text{-}wall\ tumble})=0.64$ with corr$(\Delta S,\Delta\theta)=-0.52$, against $0.45$ for the untrained baseline (\cref{fig:steer-saturation}d). If the compressive release rate is below ${\sim}10^{2}\,\text{s}^{-1}$, the wall cannot operate within the contact window, regardless of routing. Above the switching threshold, the steering is purely sign-based. The realized contact loads cluster at the ceiling (\cref{fig:steer-saturation}a, histogram), where the asymmetry is saturated, so the network cleanly resolves which side of the cell struck the wall and dictates the turn direction without relying on the magnitude of the collision.

\subsection{Behavioral performance saturates before mechanical task error}

\begin{figure}[tbp]
 \centering
  \includegraphics[width=\columnwidth]{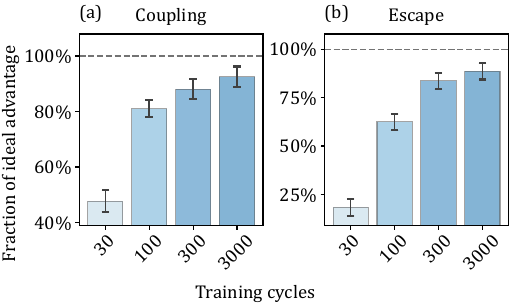}
\caption{\textbf{Behavioral gain saturates early in training.}
Networks trained for 30, 100, 300, and 3000 cycles are evaluated
in 100 independent near-wall trials per training length.
Both panels show the fraction of the ideal controller's
advantage over the untrained network recovered by training.
\textbf{(a)}~Stator--turn coupling, measured by
$\mathrm{corr}(\Delta S,\Delta\theta)$ (95\% CI).
\textbf{(b)}~First-near-wall-tumble escape probability
(Wilson 95\% CI).
The dashed line marks complete recovery of the ideal advantage;
reference values are from \cref{fig:pop-escape}.}
\label{fig:training-cycles}
\end{figure}

How accurately must the network learn the prescribed mechanical response to produce successful behavior? Because stator detachment saturates once the routed compression crosses the kinetic threshold, functional steering requires far fewer training cycles than full mechanical convergence. The canonical network is trained for 3000 cycles, over which the task error falls by more than three orders of magnitude (\cref{fig:training}a); to find how much of that reduction the behavior actually requires, we truncated training at 30, 100, and 300 cycles and evaluated each network in the same near-wall escape assay (\cref{fig:training-cycles}).

The behavioral gain is largely established by 300 cycles (\cref{fig:training-cycles}). By that point the network recovers most of the ideal steering and escape advantage, while the task error continues to fall for more than two further orders of magnitude. Once the trained pathway routes enough compression to switch the far-side occupancy faster than a contact, further reduction in readout error cannot change the motor response: the same condition that makes the steering robust to collision strength (\cref{eq:switch}) also makes it insensitive to additional training.

\subsection{Learning with crosslink turnover}
\begin{figure}[tbp]
 \centering
  \includegraphics[width=\columnwidth]{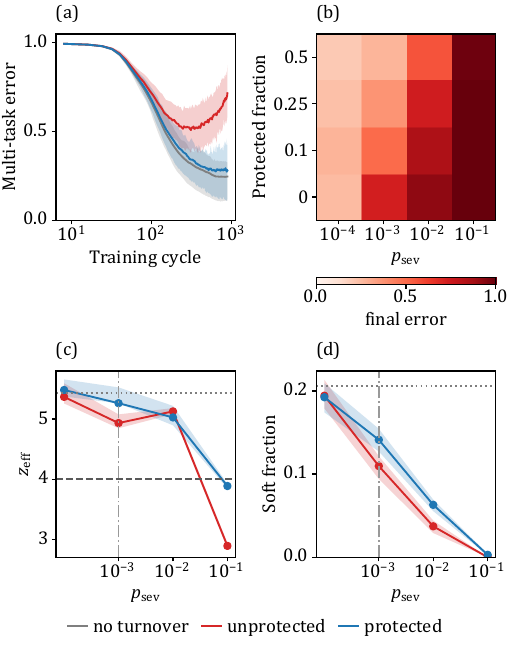}
 \caption{\textbf{Learning under peptidoglycan turnover.}
\textbf{(a)}~Multi-task error $\bar\eta_{\mathrm{RMS}}$ versus
cycle at severing probability $p_{\mathrm{sev}}=10^{-3}$:
no turnover, blind turnover ($f_p=0$), and tension-protected
turnover ($f_p=0.5$). Lines show means over 160 networks;
bands show interquartile ranges.
\textbf{(b)}~Mean final error across
$(p_{\mathrm{sev}},f_p)$; 20 networks per condition.
\textbf{(c)}~Load-bearing coordination $z_{\mathrm{eff}}$,
compared with the Maxwell estimate $z_c=4$.
\textbf{(d)}~Fraction of soft bonds
($k_{\min}<k<0.1k_0$), excluding severed bonds.
Dash-dotted lines in (c,d) mark $p_{\mathrm{sev}}=10^{-3}$.}
\label{fig:turnover}
\end{figure}

The overall architecture of the cell wall is held fixed during training in \cref{fig:training}. The changes in stiffness are due to changes in bundling and crosslinking. However, there exist transpeptidases and endopeptidases that cut and replace
crosslinks continuously~\cite{Park2008,Singh2012}. A stiffness pattern is
therefore being modified while it is simultaneously being written. Thus, we ask whether it can be
learned at all given the turnover. We add a turnover step after the stiffness
update of every cycle, following the protocol used for cytoskeletal
networks~\cite{Banerjee2026}. Each learning bond is in one of two states:

\begin{enumerate}
  \item \textbf{Intact.} Severed with probability $p_{\text{sev}}$, setting
  $k_b \to k_{\min}$. A bond whose feedforward (IN) tension magnitude $|t_b^{\rm FF}|$ exceeds a threshold
  $t_{\text{crit}}$ is protected and cannot be severed.
  \item \textbf{Severed.} Reconnected with probability $p_{\text{con}}=0.1$,
  returning the same site to the untrained stiffness $k_b \to k_0$.
\end{enumerate}

Because rest lengths equal the reference node separations and are never
trained, a reconnected bond is born unstressed. At each cycle, the threshold $t_{\text{crit}}$
is set so that the top fraction $f_p$ of intact learning bonds by tension magnitude are protected; $f_p=0$ is turnover blind
to tension and serves as the control, and performance is tracked via the root-mean-square error $\bar\eta_{\text{RMS}}$ across all eight tasks. Protection is the analogue of the
tension-stabilized bonds of Ref.~\cite{Banerjee2026} and of the load-dependent anchor lifetime in \cref{seq:koff}. It reads the feedforward state only: the feedback state
is an error field with no counterpart in a wall encounter, so we do not let it decide which bonds survive.

We calibrate the severing probability per training cycle using the frequency of wall encounters over a bacterial generation. Identifying one cycle with one wall encounter fixes $p_{\text{sev}}$ as the inverse of the number of encounters a cell experiences in one turnover time. 
Counting an encounter as a rising edge of $\max_i F_{\text{wall},i}$ through the contact gate, the trained runs give $0.22\,\text{s}^{-1}$ ($224$ onsets over $1000\,$s of simulated swimming), so a cell in continuous contact with structure sees $\sim\!400$ encounters over a $30\,$min generation. With peptidoglycan turnover on the generation timescale~\cite{Park2008}, a crosslink turns over with probability $p_{\text{sev}}\approx2.5\times10^{-3}$ per encounter; we run the calibrated condition at the nearest point on the log-spaced sweep grid, $p_{\text{sev}}=10^{-3}$.

At this biologically calibrated turnover rate, mechanical memory survives only when load-bearing bonds are shielded from severing. Specifically, turnover blind to tension destroys the learned pathway, whereas protecting the tension-bearing half restores the error to near the no-turnover value (\cref{fig:turnover}a,b). The residual difference is small on the untrained error scale and comparable to the ensemble spread. Protection acts gradually rather than as a switch, with increasing protected fraction recovering the trained error (\cref{fig:turnover}b).

The pattern does not degrade uniformly. Training drives bonds in both
directions (\cref{fig:training}b), and protection reads tension, so it shields
the stiffened channels and leaves the softened bonds exposed. The fraction of
bonds held soft by training falls from $0.21$ without turnover to $0.11$
unprotected and $0.14$ protected (\cref{fig:turnover}d). The protected network
recovers baseline error with only two-thirds of its soft complement, so the
routing does not require the full pattern, consistent with the early
behavioral saturation of \cref{fig:training-cycles}.

Failure becomes structural only at $p_{\text{sev}}=10^{-1}$, two decades above the calibrated value. There, half the learning bonds are absent at any instant and $z_{\text{eff}}$ falls to $2.9$, below the rigidity threshold that helps bound trainability (\cref{fig:turnover}c). Loss of connectivity under fast severing is also the failure mode reported for cytoskeletal networks~\cite{Banerjee2026}. These results suggest that mechanical learning can persist under continuous crosslink turnover, provided that load-bearing bonds are preferentially protected from severing. Learning must also occur on a timescale sufficiently short compared with turnover, a condition that may be favored by frequent mechanical encounters in environments such as biofilms, mucus layers, and porous media.

\section{Discussion}

Our results demonstrate how a trainable mechanical network, embedded within a model bacterium, can couple environmental contact to whole-organism behavior without requiring a centralized controller. In the model, mechanical loads generated by contact with a wall are transmitted through trained pathways in the peptidoglycan network to motor-anchor regions on the opposite side of the cell. The resulting compression increases stator detachment, producing an asymmetry in propulsive forces that biases the swimmer to turn away from the wall. On the first tumble after entrapment, the trained wall recovers 89\% of the escape advantage of an ideal side-resolved controller, despite having access only to compression-induced stator detachment. Thus, a trained structural material can serve not only as mechanical support but also as a pathway for transmitting information that influences organism-level behavior.

The effectiveness of this mechanical pathway depends not only on the transmitted signal but also on the response timescale of the downstream actuator. A routed compression that would eventually produce a large stator asymmetry cannot steer the swimmer if that asymmetry develops after contact has ended. The trained network transmits sufficient compression to produce a substantial stator response within the contact window, whereas the untrained network does not. This kinetic constraint also explains why behavioral
performance saturates before the mechanical training error is minimized: once the actuator
responds sufficiently within a collision, further improvements in mechanical transmission provide little additional behavioral benefit. More generally, the success of physical learning in an embodied system must be assessed through the dynamics of the coupled material and actuator, rather than through the accuracy of the trained mechanical response alone.

Moreover, the behavior also does not require a fully converged or static substrate. Truncating training at 300 cycles recovers $84\%$ of the ideal escape advantage, while the task error continues to fall for more than two decades (\cref{fig:training-cycles}). Under biologically calibrated turnover, blind severing destroys the pathway, but protecting the tension-bearing half of the bonds restores the error to near the no-turnover value (\cref{fig:turnover}). The protected pattern retains only part of the softened bond population, consistent with the early saturation: the behavior needs a load-bearing route that can switch the motors in time, not the full trained stiffness field. Mechanical memory can therefore persist in a wall that is continuously remodeled, provided load-bearing crosslinks are shielded and contacts occur often on the turnover timescale. The routing also requires a substrate that can support stress. In central-force networks, the onset of trainability occurs near the rigidity threshold estimated by the Maxwell criterion, $z_c=4$ (\cref{fig:sweep}). This near coincidence suggests that the connectivity required for structural rigidity also constrains the ability to establish trained mechanical pathways. However, real fiber networks can be stabilized by bending rigidity below the central-force
isostatic point~\cite{das2012redundancy}, so the trainability part here remains an open question. 

Our results complement autonomous physical learning in cytoskeletal networks~\cite{Banerjee2026}. There, actomyosin activity supplies the drive and the learned function is an internal mechanical set-point; the autonomy condition is satisfied, but the trained response does not by itself produce an organism-level action.  Here, the trained response sets a heading and changes future contact statistics, satisfying the deployment conditions with supervised training. The two requirements are separable. A learning rule that combines them would need to be driven by the agent's own encounters and to minimize a cost related to the agent's performance rather than the substrate readout error. Our update is currently applied to the wall: the collision supplies the feedforward (IN) state, but the feedback (OUT) state remains a clamped target. Deriving an online rule whose error is the escape behavior itself is the natural next step. The large separation between elastic relaxation and enzymatic remodeling places the wall in the quasistatic regime favorable for local rules~\cite{Stern2022nonequilibrium}; sequential presentation of tasks would still need to avoid interference~\cite{Stern2020continual}.

In mechanically structured habitats, this pathway would operate alongside chemotaxis. Chemical gradients guide cells over seconds, but porous soil, mucus, biofilms, and crowded colonies repeatedly impose contact on the cell body. In those settings, the immediate problem is often how to reorient after contact, and the relevant timescale is the collision. Interestingly, recent studies have further demonstrated load-dependent mechanoadaptation and mechanically mediated cooperative switching in the \Ecoli{} flagellar motor~\cite{Wadhwa2022,Mattingly2026}, suggesting additional ways in which mechanical signals can influence bacterial motility.

Given our results, two experimental tests follow. First, repeated wall contact should bias peptidoglycan remodeling near motor-anchor regions if the wall adapts to contact history. Second, cells with a history of controlled wall contact should escape near-surface trapping sooner than naive cells; a simple assay would flow motile \Ecoli{} through shallow posts or constrictions and then measure escape statistics in a fresh chamber. The kinetic hypothesis gives a further test: steering through a sequence of alternating left- and right-side contacts (such as a staggered post array) would require stator reloading on the timescale of the interval between contacts, which is faster than the baseline on-rate used here. 

Three limitations bound these claims. First, the network is much smaller than a peptidoglycan sacculus. Long-range routing may become harder at biological size, and whether the same transfer ratio survives in a full sacculus remains open~\cite{Huang2008,Rocks2017}. Second, the compression-release channel is a hypothesis. No experiment has tested whether compressing a peptide crosslink below the MotB binding span increases stator detachment, and the prefactor $k_{\mathrm{off}}^{\mathrm{geo}}$ is a modeling choice constrained by the requirement that the switch occur within a contact (\cref{eq:switch}; S4.3). The network itself is also a linear, unpressurized central-force model. A real sacculus is prestressed by turgor and is often modeled with nonlinear chain elasticity~\cite{Nguyen2015}; we used the unstressed linear network because it is the minimal substrate in which stiffness routing can be isolated.  Prestress rigidifies a sub-isostatic central-force network, so turgor pressure could place a sacculus above the effective threshold at a coordination where our unstressed model fails. If that is the case, the condition on the biological substrate is a condition on both the crosslink density and turgor pressure, and the trained routing should degrade when the cell is osmotically deflated. Third, we must move beyond quasistatic training to dynamic training to uncover the full potential for behavioral adaptation. 

For now, the broader question is whether structural mechanics can serve as a control pathway in cells. Recent work demonstrates that training a packings of cells to achieve a target stress pattern via structural mechanics, though the target is a trainer-prescribed one independent of a specific biological function~\cite{ameen2026training}. If a load-bearing material can be tuned to bias an actuator during the relevant stimulus window, it can participate in behavior without an additional intracellular signaling cascade. We do not claim that the present model demonstrates cell-scale cognition; rather, it represents a step toward satisfying all four conditions for autonomous embodied physical learning. However, we show one trained behavior in a minimal system and a condition that such a behavior must satisfy: the trained substrate must deliver enough signal for the actuator to complete its response while the stimulus is present. A readout of the network alone would not show whether this condition is met. Thus, such a framework can be extended to other types of cells, such as slime mold where local adaptation rules produce functional
global responses without externally prescribed
training signals~\cite{boussard2021adaptive,alim2026decision, anisetti2023slime}.

\section*{Data and code availability}
All code and data supporting the results will be made available upon publication.

\section*{Acknowledgements}
We thank Vidyesh Anisetti, Nidhi Pashine, Sadjad Arzash and Abhinandan Aman for helpful discussions. JMS acknowledges the National Science Foundation under DMR-2204312.

\section*{Competing interests}
The authors declare no competing interests.
\bibliography{refs}
\clearpage

\section*{Supplementary Material}
\setcounter{figure}{0}
\renewcommand{\thefigure}{S\arabic{figure}}
\setcounter{equation}{0}
\renewcommand{\theequation}{S\arabic{equation}}
\setcounter{table}{0}
\renewcommand{\thetable}{S\Roman{table}}

\section*{S1. Disordered spring network}
The mechanical cell wall is represented as a two-dimensional disordered spring network. To encode the disorder, node positions are generated by quenching a bidisperse mixture (radius ratio 1.4, effective packing fraction $\phi = 0.84$) to its local energy minimum (inherent structure). The particles interact via a dimensionless 12--6 Lennard-Jones potential, $U_{ij} = (\sigma_{ij}/r_{ij})^{12} - (\sigma_{ij}/r_{ij})^6$, with contact scales $\sigma_{ij} = r_i + r_j$, minimized over all pairs in a square box $L_x = L_y = 10$ (lattice units) with periodic boundary conditions (a torus). After the node positions are established, two nodes are connected by a Hookean spring when their scaled separation $r_{ij}/\sigma_{ij}$ is below a connection cutoff of 1.3. An optional dilution probability 1-$p$ removes a fraction of bonds (used only for the connectivity sweep, S3). For the trained network, $N = 512$ nodes and $N_b = 1497$ bonds, so the mean coordination is $z = 2N_b/N \approx 5.85$.

As for the mechanics of the network, each bond $\langle ij\rangle$ carries a stiffness $k_{ij}$ and a rest length
$\ell^0_{ij}$. The stiffness is the only trainable degree of freedom; rest
lengths are held fixed.
The choice of learning degree of freedom determines what class of input--output
map the trained network can express. Rest-length training displaces the
source--target relation, shifting its intercept and leaving the network
pre-strained, while stiffness training alters the slope with the network
unstrained at rest; neither alone spans a general affine map, and representing
one requires both~\cite{Banerjee2026}. The task here is a fixed transfer ratio
between an applied anchor load and a far-side compressive readout, a
slope. A rest-length solution would satisfy that ratio only at the trained input
magnitude and would carry a standing pre-strain in its absence, so the response
would not extend across the range of contact loads the cell encounters, nor
vanish when the cell is in open field. Stiffness training gives a
scale-free transfer ratio (S6), the property the near-wall assay
requires. It is also the only choice compatible with the turnover protocol (\cref{fig:turnover}),
in which a reconnected bond is born unstressed: a pre-strain pattern would be
erased by each reconnection, whereas a stiffness pattern is not. All stiffnesses are initialized uniformly at $k_0 =
10^4$, so the disorder that training exploits comes from the network
topology, not from any stiffness randomization. Because both training states prescribe tensions, the tension field is invariant under a uniform rescaling of all stiffnesses, so $k_0$ sets no quantity we report; it fixes the displacement scale, and with it whether the linear-response solve is a valid description. Stiffnesses are clipped to $[k_{\min}, k_{\max}] = [10^{-10}, 10^{10}]$, i.e.\ effectively unclamped, so
trained bonds can reach both very high and very low stiffness; this is why the
trained $\log(k/k_0)$ map spans roughly $\pm 6$.

To achieve mechanical equilibrium, the stretching energy
\begin{equation}
  E = \sum_{\langle ij\rangle} \tfrac12\, k_{ij}
  \bigl(|\vect{r}_i - \vect{r}_j| - \ell^0_{ij}\bigr)^2,
  \label{seq:energy}
\end{equation}
is minimized subject to the imposed boundary conditions. Thus,
the cell wall is treated quasistatically, without explicit
dynamics. To compute its mechanical response, we linearize
the elastic energy about the reference configuration and solve
the resulting boundary-value problem using sparse factorization.

The quasistatic approximation is motivated by the separation
between elastic relaxation and contact timescales. For overdamped
node dynamics,
$\zeta\,\dd\vect{r}/\dd t=-\partial E/\partial\vect{r}$,
a single bond relaxes on a timescale
$\tau_{\text{bond}}=\zeta/k$. Assuming diffusive relaxation
of long-wavelength elastic modes, the longest network relaxation
time scales as
$\tau_{\text{net}}\sim\tau_{\text{bond}}(L/a)^2$.
Using the Stokes drag on a nanometre-scale node in water and an
effective peptide stiffness estimated from the MD-derived
force--extension model of Ref.~\cite{Nguyen2015}, we obtain
$\tau_{\text{bond}}\sim1\,\mathrm{ns}$. This estimate uses
physical parameters rather than the dimensionless parameters
of our simulations and neglects additional sources of
dissipation within the cell wall. For a patch comparable in
size to our network ($L/a\sim20$), it gives
$\tau_{\text{net}}\sim1\,\mu\mathrm{s}$, approximately five
orders of magnitude shorter than the median contact duration
($0.11\,\mathrm{s}$, S5.1). Extrapolating the same scaling
to the full cell wall ($L/a\sim10^3$) gives
$\tau_{\text{net}}\sim1\,\mathrm{ms}$, still shorter than
the contact duration, although by a smaller margin.
These estimates do not account for anomalously slow modes
that may arise in disordered networks, nor do they directly
determine the relaxation time of the trained input--output
response. When elastic relaxation becomes comparable to the
contact duration, the mechanical signal may not reach its
quasistatic value before contact ends. Finite relaxation
could therefore reduce the effectiveness of steering during
short encounters, introducing an additional dynamical
constraint beyond the stator kinetics considered here.

estimate gives $\tau_{\text{net}}\sim1\,$ms, shorter than a contact, but by a smaller margin. Where the separation fails, the routed signal arrives later than the solve assumes. The escape statistics are therefore an upper bound, while the condition of \cref{eq:switch} binds harder than we report: elastic delay adds to $\tau_S$ without lengthening the contact. 

Motors are mapped from the rod-shaped cell onto the sheet by unrolling a
cylinder: the cell long axis maps to the network $y$ coordinate, and a motor's
left/right side (sign of its lateral position on the cell) maps to one of the two
seams $x = 0$ or $x = L_x/2$. Each motor is associated with a center node and a
ring of nearby edges (the stator-anchor locus); within that ring a single
input edge (where wall force enters) and a single readout edge
(where the trained tension is measured) are assigned, with these input/output
assignments frozen once chosen.

\section*{S2. Training: Multi-mechanism learning}

The stiffness pattern is trained by a local, two-state update, a variation on the multi-mechanism learning rule~\cite{Anisetti2023} (MmL) that compares two static
mechanical experiments on the network each cycle:
\begin{enumerate}
    \item \textbf{Feedforward (IN) state.} Input edges are driven under displacement control, with the
  imposed strain $\varepsilon = t_{\text{pin}}/(k\,\ell^0)$ chosen so that the edge carries
  the target tension $t_{\text{pin}}$ at its current stiffness; $\varepsilon$ is therefore
  recomputed each cycle as $k$ evolves. The linear response is solved and the resulting
  signed bond tension $t^{\rm FF}_b$ is recorded on every bond $b$.
  \item \textbf{Feedback (OUT) state.} Linear response with clamped tensions imposed on the
  target (output) edges and no input forces, recording
  $t^{\rm FB}_b$.
\end{enumerate}

The clamped target tension on each output edge is set by a proportional rule,
\begin{equation}
    t^{\rm FB}_e = t^{D}_e - t^{\rm F}_e, \qquad e \in \mathcal{T},
  \label{seq:proportional}
\end{equation}
where $t^{D}_e$ is the desired tension and $t^{\rm F}_e$ the current (free) tension measured in the feedforward state. The
stiffness of every learning bond is then updated by the signed product of its two
recorded tensions,
\begin{equation}
  \Delta k_b = -\gamma_l\, t^{\rm FF}_b\, t^{\rm FB}_b,
  \qquad k_b \leftarrow k_b + \Delta k_b,
  \label{seq:mlink}
\end{equation}
with learning rate $\gamma_l$, additive update, and clipping to $[k_{\min},
k_{\max}]$; rest lengths are unchanged. This is the fixed-geometry form of the
vector dot product $\vect{t}^{\rm FF}_b\!\cdot\!\vect{t}^{\rm FB}_b$
implemented in the code, the two coinciding because the bond directions $\hat{\vect{n}}_b$ are evaluated at the same reference geometry in both states. \cref{seq:mlink} is the mechanical analogue of the non-interfering feedback rule
of Anisetti \textit{et al.}~\cite{Anisetti2023}, whose flow-network update
$\Delta w_{xy} = -\alpha\,u(x,y)\,v(x,y)$ multiplies two local drops across the
same edge---one carrying the activation signal, one the error---each a steady
state of the same Laplace operator under different boundary conditions. Here the
two signals are the IN- and OUT-state tensions on the same bond, each a linear
response of the same stiffness matrix. 

\begin{figure}[tbp]
 \centering
  \includegraphics[width=\columnwidth]{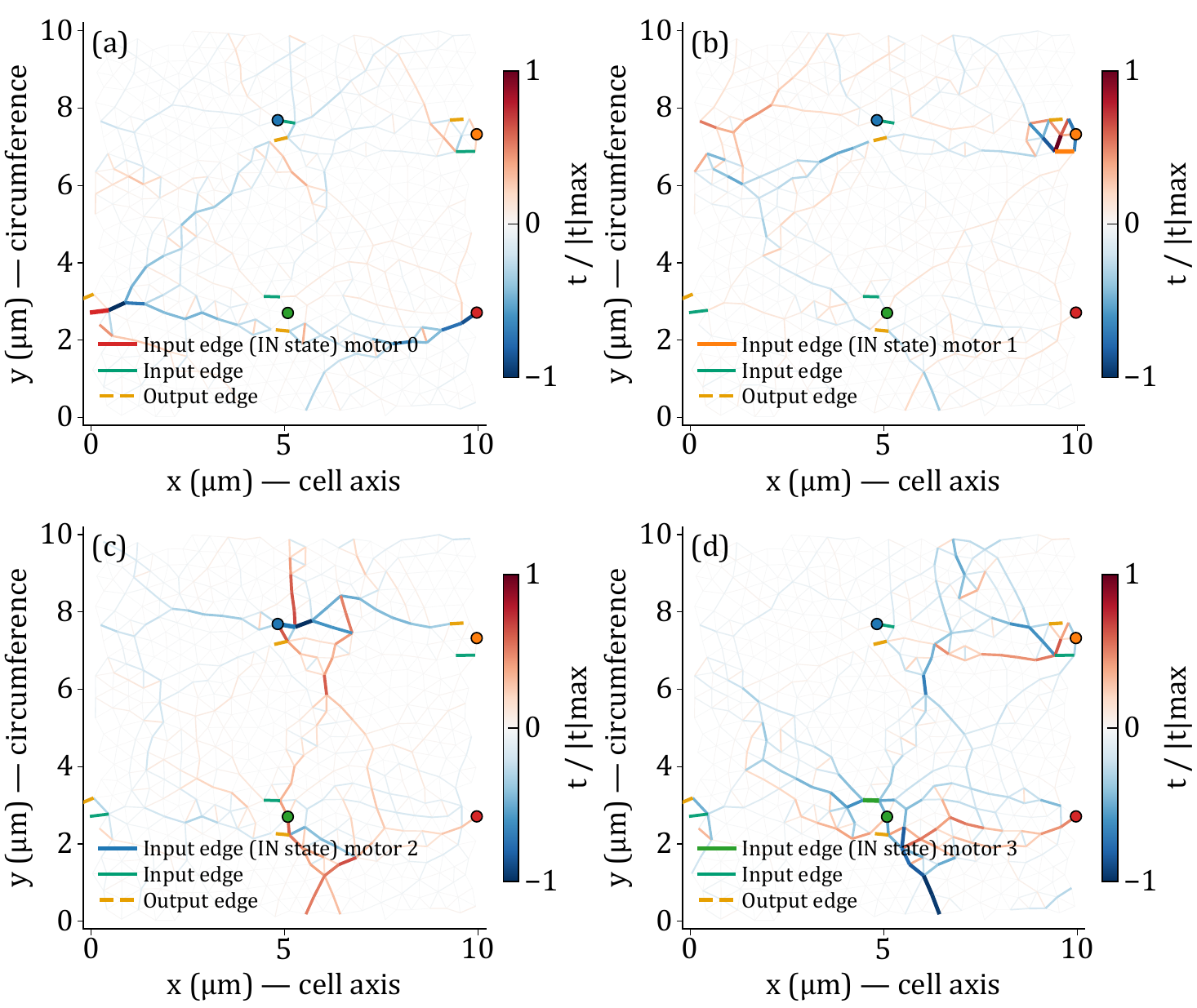}
  \caption{\textbf{IN-state mechanical response across eight cyclic training tasks.}
The network is trained on eight input--output pairings: each task designates one
motor as impacted (input load) and one motor on the opposite side as the
non-impacted readout (output). Panels \textbf{(a)}--\textbf{(d)} show normalized
IN-state bond tensions $t/|t|_{\max}$ after training when the input load is applied
at motor 1, 2, 3, and 4 respectively (remaining tasks rotate the input/output
assignment around the four sites). Applied load no longer spreads uniformly across the network;
instead, stiffened bonds form contiguous chains that connect the loaded anchor to
the designated far-side readout. These maps are the mechanical substrate for the
population routing and steering statistics in \cref{fig:pop-escape,fig:steer-saturation}.}
  \label{fig:si_tension_panels}
\end{figure}

Training quality is monitored by the normalized readout error
\begin{equation}
  \eta = \frac{1}{t_0}\sqrt{\frac{1}{N_T}\sum_{e\in\mathcal{T}}\bigl(t^{D}_e - t^{F}_e\bigr)^2},
  \label{seq:error}
\end{equation}
over the $N_T$ target edges $\mathcal{T}$, normalized by the target tension scale $t_0 = |t^D_e|$. To show how the response depends on the loaded motor,
\cref{fig:si_tension_panels} presents IN-state tension maps
for input loads applied to each of the four motors in turn.

The canonical run trains eight input--output tasks on a cyclic schedule (one task per cycle), with learning rate $\gamma_l = 1.11\times10^{8}$, equivalently $\Gamma \equiv \gamma_l\,t_{\text{pin}}^{2}/k_0 = 10^{3}$ in the dimensionless step of \cref{seq:mlink}, over $n_{\text{cyc}} = 3000$ cycles with convergence threshold $10^{-3}$. The full network and training
parameters are listed in \cref{tab:net-train-params}.

\section*{S3. Connectivity and the rigidity transition}

To systematically probe the dependence of trainability on network rigidity, we train the network across a sweep of bond-dilution probabilities $p$ (from 0.3 to 1.0 in 10 increments). For each probability, we average over 100 independent network instances across system sizes $N \in \{64, 256, 512\}$. Decreasing $p$ reduces the mean coordination $z$ toward and below the Maxwell isostatic point of $z_c = 4$, the theoretical rigidity threshold for central-force networks~\cite{Maxwell1864, jacobs1995generic}. For each instance, we compare the trained task error against the untrained baseline, and evaluate the floppy-mode fraction $N_{\text{fl}}/N_{\text{DOF}}$ against the Maxwell counting estimate of $1 - z/z_c$. While trainability in our purely central-force model emerges near the Maxwell mean-field estimate of $z_c = 2d = 4$, real biological networks often operate in a sub-isostatic regime ($z < 4$), where bending rigidity and macroscopic strain-stiffening provide alternative mechanisms for mechanical stability~\cite{das2012redundancy, Broedersz2014,Sharma2016}.
\section*{S4. Agent: motor dynamics and motility}

\subsection*{S4.1 Receptor activity in a uniform field}

The agent senses ligand through the log-concentration of the field convolved with a biphasic memory kernel~\cite{Tu2008}. We utilize the specific discrete biphasic kernel approximation formalized by Chatterjee \textit{et al.}~\cite{Chatterjee2011}, whose integral is zero by construction. Consequently, the response adapts exactly in a constant field~\cite{Shoval2010,Yi2000,Chatterjee2011}. Because the wall experiment uses a spatially uniform field, the sensing mechanism returns a constant signal and the receptor activity is held at its baseline $H = h_0 = 1.3$, leaving run--tumble switching (S4.2) as the only active navigation behavior.

\subsection*{S4.2 Run--tumble switching}

\begin{figure}[tbp]
 \centering
  \includegraphics[width=\columnwidth]{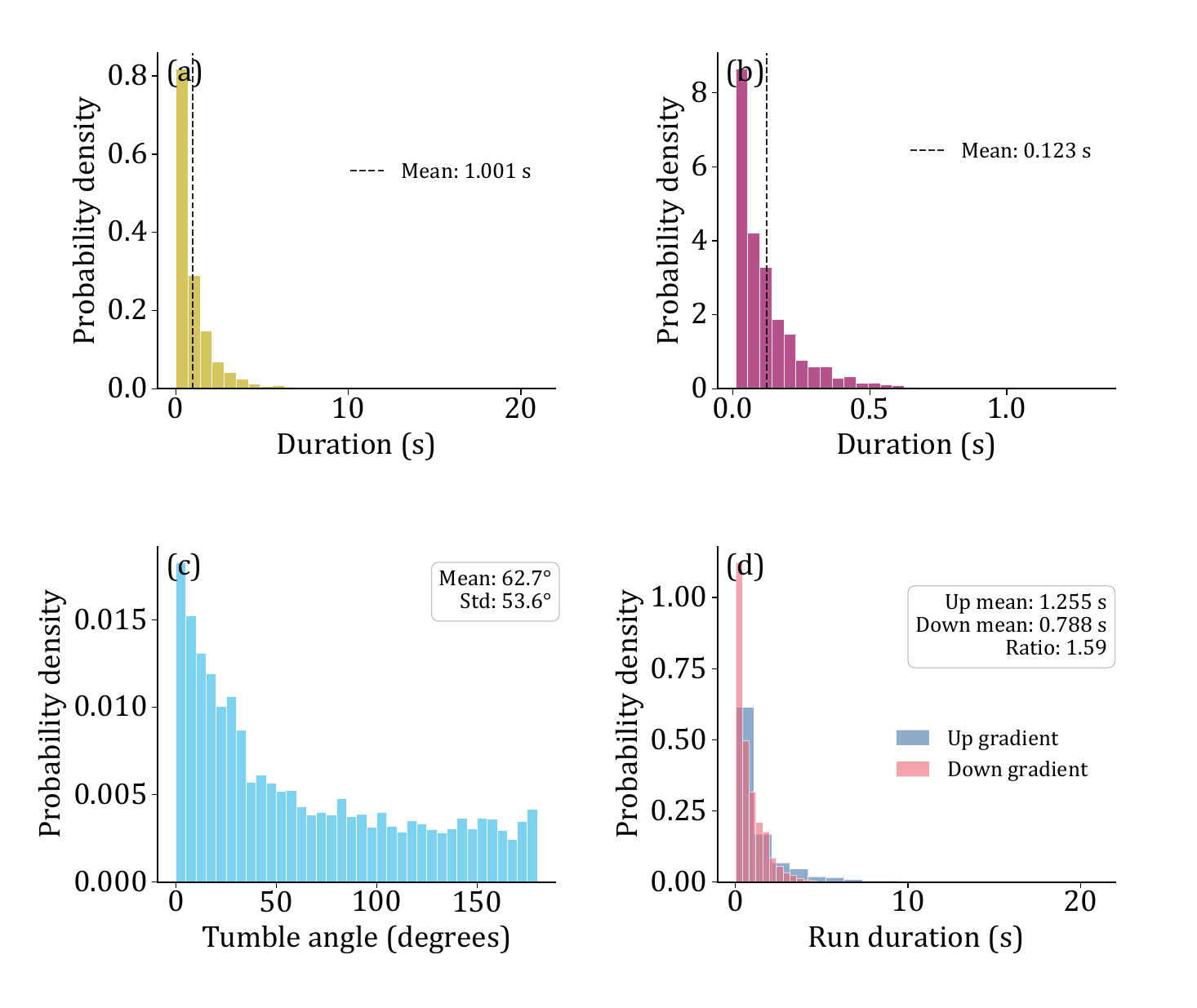}
  \caption{\textbf{Baseline run--tumble calibration.}
Distributions of run and tumble durations produced by the stochastic switching
rule \cref{seq:switching} at baseline receptor activity $H=h_0=1.3$ with Hill
exponent $h=2.5$, $K_{1/2}=1.25$, $k_{+}=1.66\,\text{s}^{-1}$, and
$k_{-}=11.6\,\text{s}^{-1}$. Mean tumble frequency
$\langle f_{\text{tum}}\rangle=0.87\,$Hz and mean tumble duration
$0.12\,$s match canonical \Ecoli{} values ($\sim1\,$Hz, $\sim0.1\,$s)~\cite{Berg1972}.
The resulting mean run time $\langle\tau_{\text{run}}\rangle\approx1.0\,$s sets
the trap escape criterion \cref{seq:escape} used in the near-wall assay.}
  \label{fig:calibration}
\end{figure}

The agent is a rod-shaped swimmer with $N_{\text{mot}} = 4$ flagellar motors. Each motor switches stochastically between counter-clockwise (CCW, run) and
clockwise (CW, tumble) rotation. The per-motor switching rates are Hill functions
of the receptor activity $H$,
\begin{equation}
  k_{\mathrm{CCW}\to\mathrm{CW}} = k_{+}\frac{H^{h}}{H^{h}+K_{1/2}^{h}}, \qquad
  k_{\mathrm{CW}\to\mathrm{CCW}} = k_{-}\frac{K_{1/2}^{h}}{H^{h}+K_{1/2}^{h}},
  \label{seq:switching}
\end{equation}
where $H$ is the dimensionless receptor activity of S4.1 (not a CheY-P
concentration) and $K_{1/2} = 1.25$ is on the same dimensionless scale. The Hill
form for ultrasensitive motor switching as a function of receptor state is
established~\cite{Cluzel2000,Yuan2013}. The specific values used here,
$h = 2.5$, $K_{1/2} = 1.25$, $k_{+} = 1.66\,\text{s}^{-1}$, and $k_{-} =
11.6\,\text{s}^{-1}$, are not taken from those measurements; they are calibration
constants chosen so that, at the baseline activity $H = h_0$, the cell reproduces
the canonical run--tumble statistics of \Ecoli{} ($\sim 1\,$Hz tumble frequency,
$\sim 0.1\,$s tumble duration)~\cite{Berg1972}. The calibration is shown in
\cref{fig:calibration}. Each motor flips with probability $k\,\Delta t$ per step
($\Delta t = 0.01\,$s) against a uniform draw, and the cell is declared tumbling
when at least $N_{\text{th}} = 2$ of the $4$ motors are
CW. We adopt a coarser threshold $N_{\text{th}} = 2$ so that a single transient CW motor does not trigger a tumble---a simplification of the correlation mechanism of~\cite{Mears2014}.

\subsection*{S4.3 Stator occupancy and the load-dependent off-rate}

The occupancy dynamics of \cref{seq:stator} and the two-channel off-rate of \cref{seq:koff} are introduced in the Model; here we give the parameter values and their basis. We take $N_{\text{stat}} = 11$~\cite{Reid2006,Leake2006}, $k_{\text{on}} = 0.01\,\text{s}^{-1}$, $k_{\text{off}}^{\text{LS}} = 0.057\,\text{s}^{-1}$, $k_{\text{off}}^{\text{geo}} = 600\,\text{s}^{-1}$, $\delta_\text{LS} = 2.48\,$nm, $\delta_{\text{geo}} = 1.5\,$nm, and $\beta = 0.24\,(\text{pN}\cdot\text{nm})^{-1}$. 

Holding the on-rate constant and placing all load-dependence in $k_{\text{off}}$ follows Wadhwa \textit{et al.}~\cite{Wadhwa2021}, who fit $k_{+}=0.0016\,\text{s}^{-1}$ across their electrorotation range, where $k_{+}$ is the on-rate in our notation, and report a larger on-rate during high-torque recovery without quantifying it. Our motors run near stall (S4.4), so we take a value in that elevated regime; a slower on-rate would lengthen the recovery time and lower the saturation threshold of \cref{eq:switch}, strengthening both kinetic conclusions of the Results.

The first term is the load-stabilizing channel~\cite{Nord2017,Chawla2017}: a larger per-stator anchor force $F_{\text{stat},i}$ lowers the detachment rate, so stators engage longer
under load. The baseline rate is taken from the load-dependent off-rate
of~\cite{Wadhwa2022} and the characteristic distance from~\cite{Chawla2017}. The
anchor force is itself load-dependent (S4.4), maximal at stall and vanishing at the
free-running speed. We model only the load-stabilizing (catch) branch of the stator anchor bond. A
full catch--slip bond superposes a force-weakened catch channel and a
force-strengthened slip channel, giving a lifetime that peaks at a nonzero
force~\cite{Nord2017}; we retain only the load-stabilizing exponential, in the
single-exponential form fit by Chawla \textit{et al.}~\cite{Chawla2017}. This is
appropriate because the anchor force stays at or below the per-stator stall value
$\tau_{\text{stat}}/r_{\text{mot}}\approx 11\,$pN (\cref{seq:fstator}), the
load regime over which the interface behaves as a catch bond;
Nord \textit{et al.}~\cite{Nord2017} predict the crossover to slip only above the
maximum force a single unit generates. Importantly, the second term of
\cref{seq:koff} is not the slip branch of this bond but a distinct,
hypothesized geometric detachment channel, described next. 

The second term carries the trained network's signal and is the destabilizing
regime of the same geometric anchoring mechanism (Model): when the trained network
drives the local crosslink into compression ($\sigma_{\text{PG},i}<0$), the
geometric match for the MotB span is broken and the off-rate rises. The
$\exp(\cdot)-1$ form makes this channel vanish identically in the absence of
compression, so it acts only when the readout edge is compressed and leaves the resting occupancy unperturbed.

The baseline rate $k_{\text{off}}^{\text{geo}}$ and
the compression distance $\delta_{\text{geo}}$ are modeling choices (\cref{tab:bio-params}). No experiment constrains either; $k_{\text{off}}^{\text{geo}}$ is required to be $\gtrsim10^2\,\text{s}^{-1}$ so the anchor switches within a contact at physiological load (the condition \cref{eq:switch} of the Results). This requirement sits comfortably within the standard $10^{2}$--$10^{4}\,\text{s}^{-1}$ range for protein unbinding attempt rates.

Equation~\eqref{seq:stator} is integrated by explicit Euler with $\Delta t = 0.01\,$s. At the kinematic ceiling $F_{\text{wall}} = 0.3\,$pN and the measured transfer
ratio $0.905$, the geometric channel reaches $k_{\text{off}} \approx 62\,\text{s}^{-1}$, so $k_{\text{off}}\Delta t < 1$ throughout and no quasi-steady-state substitution is needed. The relaxation time under load, $\tau_S = 1/(k_{\text{on}}+k_{\text{off}}) \approx 16\,$ms, is short compared with the median contact duration ($0.11\,$s, S5.1) but not compared with $\Delta t$, so the transient is resolved rather than slaved. The stator number is initialized at its open-field steady state, evaluated with the
geometric channel off. The load stabilizing-bond rate depends on $F_{\text{stat}}$, which
depends on $S_i$ through \cref{seq:omega,seq:fstator}, so this is a fixed point,
\begin{equation}
  S^{(0)} = \frac{N_{\text{stat}}\,k_{\text{on}}}
  {k_{\text{on}} + k_{\text{off}}^{\text{LS}}
   \exp\!\bigl(-\beta F_{\text{stat}}(S^{(0)})\,\delta_\text{LS}\bigr)},
  \label{seq:S0}
\end{equation}
which iterates to $S^{(0)}=10.70$, $\omega=361\,$rad/s, $F_{\text{stat}}=9.0\,$pN.

\subsection*{S4.4 Torque, speed, and post-tumble heading}
\begin{figure}[tbp]
 \centering
  \includegraphics[width=\columnwidth]{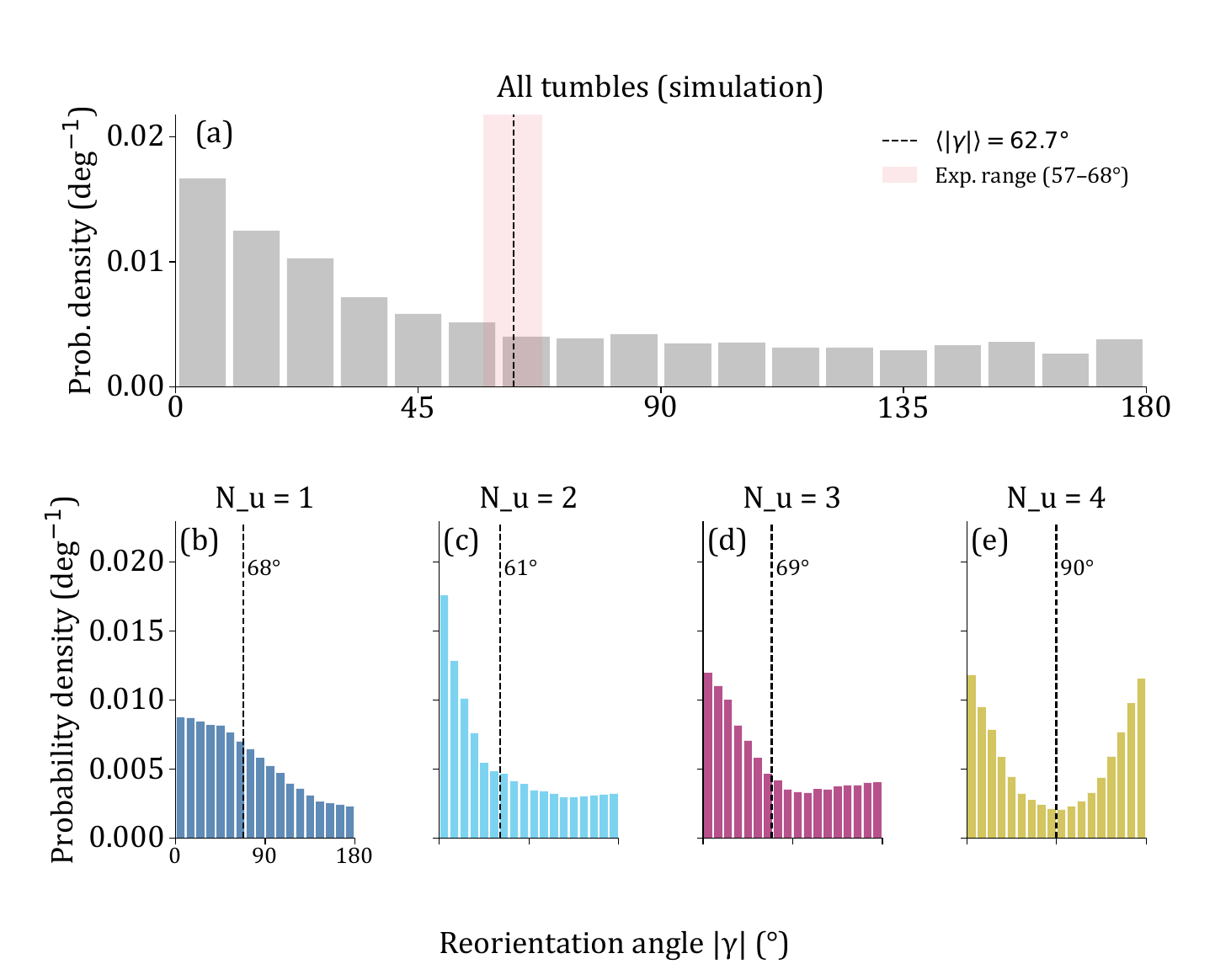}
  \caption{\textbf{Post-tumble reorientation angle decomposed by CW motor count $N_u$.}
\textbf{(a)}~Distribution of reorientation angle $\gamma$ over all simulated
tumbles; mean (dashed) lies in the experimental range 57--68$^\circ$ (shaded).
\textbf{(b)--(e)}~Same distribution conditioned on $N_u\in\{1,2,3,4\}$ simultaneously
clockwise motors. The sharp peak at $0^\circ$ arises from transverse-force
cancellation between opposite-side CW motors and is strongest at $N_u=2$ (panel~c),
the dominant case. At $N_u=4$ (panel~e) no bundled motor survives to break fore--aft
symmetry, yielding a bimodal distribution about $0^\circ$ and $180^\circ$. This
decomposition explains why the planar forced-sphere reduction peaks near the cell
axis yet retains a mean angle consistent with experiment.}
  \label{fig:tumble-decomp}
\end{figure}
The wall force acts on the motors only through $\sigma_{\text{PG}}$ in
\cref{seq:koff}; it does not enter the torque balance. The rotation speed follows
from the stator-driven balance
\begin{equation}
  \omega_i = \frac{S_i\,\tau_{\text{stat}}}{\zeta_r + S_i\,\tau_{\text{stat}}/\omega_{\max}} \le \omega_{\max},
  \label{seq:omega}
\end{equation}
with $\tau_{\text{stat}} = 250\,$pN$\cdot$nm~\cite{Wang2022}, 
$r_{\text{mot}} = 22.5\,$nm, $\zeta_r = 6.0\,$pN$\cdot$nm$\cdot$s, and
$\omega_{\max} = 1885\,$rad/s. The per-stator anchor force entering the load stabilizing-bond
channel of \cref{seq:koff} is load-dependent,
\begin{equation}
  F_{\text{stat},i} = \frac{\tau_{\text{stat}}}{r_{\text{mot}}}\left(1 - \frac{\omega_i}{\omega_{\max}}\right),
  \label{seq:fstator}
\end{equation}
maximal at stall ($\omega_i\to0$) and vanishing at the free-running speed; it is
evaluated each step from the current $\omega_i$ (hence from the current $S_i$ via
\cref{seq:omega}) before the occupancy is advanced.

The post-tumble heading is set once, at the end of the tumble, as the direction
of the vector sum of propulsive forces from the flagella, following the
forced-sphere reduction of Dvoriashyna and Lauga~\cite{Dvoriashyna2021}. In that
model the post-tumble swimming direction is the direction of the net propulsive
force, with each flagellar force pointing toward the cell body. We adopt the same
force-summation picture in the plane: because the swimmer is two-dimensional,
each motor's force is confined to the body-ward half-plane (upper motors
$U(-\pi,0)$, lower motors $U(0,\pi)$), the planar image of the toward-body constraint of~\cite{Dvoriashyna2021}, which prevents the flagellum from intersecting the cell body. The bundled (CCW) flagella contribute a force along the cell axis, and
\begin{equation}
  \vect{F}_{\text{tot}} = f_{\text{bun}}\,\hat{\vect{e}}_x
  + \sum_{i\in\text{CW}} f_i\,\hat{\vect{n}}_i,
  \qquad \gamma = \operatorname{atan2}(F_{\text{tot},y}, F_{\text{tot},x}),
\end{equation}
where each $f_i$ is the resistive-force-theory propulsive force (S4.5) and
 $\gamma$ is the post-tumble heading. Rotational drift during the tumble (often modeled as rotational diffusion~\cite{Saragosti2012}) is neglected.

This planar reduction does not reproduce the angular distribution
of~\cite{Dvoriashyna2021}, whose three-dimensional model samples the flagellar
forces isotropically over the sphere and yields a distribution that rises from
zero to a peak at intermediate angles ($\approx 40$--$50^\circ$). In the plane,
the body-ward half-plane constraint places motors on opposite sides of the body
into opposing half-planes, whose transverse force components partially cancel
when those motors are simultaneously CW; we attribute the resulting concentration of $\gamma$ along the cell axis---and, when a forward bundle survives, at small angles in particular---to this cancellation (\cref{fig:tumble-decomp}), which isotropic three-dimensional sampling averages away. The mean reorientation,
$\langle|\gamma|\rangle = 62.7^\circ$ (\cref{fig:tumble-decomp}), nonetheless falls
within the experimentally reported range ($\approx 57$--$68^\circ$)~\cite{Berg1972,Dvoriashyna2021}.
\paragraph{\Cref{fig:tumble-decomp}.}
Panel~(a) is drawn directly from the calibration simulation
(\cref{fig:calibration}; $n=4{,}446$ tumbles). Panels~(b)--(e) are
generated by Monte Carlo from the same force-summation model: for each $N_u
\in \{1,2,3,4\}$ the $\binom{4}{N_u}$ motor-assignment combinations are
sampled with equal weight ($4\times10^4$ total draws per $N_u$, spread
evenly across combinations). The bundle is treated as a single effective
filament at the mean stator occupancy of the calibration run
($\bar{S}=10.70$, giving $\omega=361\,$rad/s from \cref{seq:omega}), matching the in-simulation
draw. The panels decompose the distribution to isolate why panel~(a) peaks at $0^\circ$ and not at the $\approx40$--$50^\circ$ peak of the three-dimensional model: the $0^\circ$ peak requires two opposite-side CW motors whose transverse forces cancel, so it is sharpest at $N_u=2$ (panel~c), absent at $N_u=1$ (panel~b, no opposite partner), and split between $0^\circ$ and $180^\circ$ at $N_u=4$ (panel~e, where no bundle survives to break the fore--aft symmetry). Because $N_u=2$ dominates, the overall distribution inherits its sharp $0^\circ$ peak, and the flat tail of that same distribution lifts the mean into the experimental band.

\subsection*{S4.5 Propulsion and drag}

Each flagellum's propulsive force uses resistive-force-theory
coefficients~\cite{Purcell1977, Chattopadhyay2006, Lauga2009}:
\begin{align}
  f_i &= (\zeta_\perp - \zeta_\parallel)\,\sin\beta_i\cos\beta_i\,R_i\,L\,\omega_i,
  \nonumber\\
  \zeta_\parallel &= \frac{2\pi\mu}{\ln(L/h_f) - 0.5}, \qquad
  \zeta_\perp = \frac{4\pi\mu}{\ln(L/h_f) + 0.5},
\end{align}
where $\zeta_\parallel$ and $\zeta_\perp$ are the tangential and normal drag coefficients per unit length, $\mu$ is the fluid viscosity, and $L$, $h_f$, $R_i$, and $\beta_i$ are the filament contour length, half-thickness, helix radius, and pitch angle, respectively (\cref{tab:bio-params}). At the free-running occupancy of the calibration run, the normal-helix parameters give $f_i \approx 0.2\,$pN per filament, and the semi-coiled geometry gives $\approx 0.3\,$pN. Stokes drag on the cell body at $v_{\text{run}}$ is of the same order. This is the physiological thrust scale referred to in the Results.

\section*{S5. Near-wall simulation: contact loop, trapping, and the escape statistic}
\subsection*{S5.1 Wall contact}
The wall experiment couples the agent to the trained network one step at a time.
The arena is a horizontal channel of height $100\,\mu$m and width $500\,\mu$m,
periodic in $x$ and hard-walled in $y$, with wall thickness $5\,\mu$m. The
standard run integrates $100{,}000$ steps at $\Delta t = 0.01\,$s ($1000\,$s) and
averages over $n_{\text{sim}}=100$ independent trials.

At each step:
\begin{enumerate}
  \item The wall contact force on each motor is a Hookean penetration term,
  $F_{\text{wall},i} = k_{\text{wall}}(d_{\text{buf}} - d_i)\,
  |\hat{\vect{n}}^{\text{flag}}_i\!\cdot\!\hat{\vect{n}}^{\text{wall}}|$,
  nonzero only when the tip of flagellum $i$ lies within the buffer distance
  $d_{\text{buf}}$ of a wall face. Here, $d_i$ is the tip--wall distance, and the
  projection onto the wall normal ensures a glancing flagellum registers less impact than a
  perpendicular one. We clamp the tip--wall distance at $d_i \ge 0$ (no wall penetration), so the signal saturates at $k_{\text{wall}}d_{\text{buf}} = 0.3\,$pN.
\item If the maximum wall force across motors is below the contact threshold ($0.01\,$pN), the cell is in open field: the PG step is skipped and $\sigma_{\text{PG}} \to 0$. Otherwise the single designated input edge of each contacting motor (S1) is pinned
at a tension $t^{\text{in}}_i = c\,F_{\text{wall},i}$ with $c = 1$, so the pinned
tension equals the contact force in magnitude. This is a modeling choice: the
contact force is delivered to one crosslink of the anchor locus rather than distributed across the ring. The network is equilibrated (S1),
  and the tension on each motor's readout edge is read out as $\sigma_{\text{PG},i}$
  and fed to \cref{seq:koff}.
  \item The agent advances (S4): switching, stator update, motion.
\end{enumerate}

The contact term acts as a geometric sensor rather than a true hydrodynamic load. Run motion is prescribed at $v_{\text{run}}$ and a step that would place the cell body inside a wall is simply rejected. Consequently, the body is hard-walled at its own radius and never sits in physical force-balance against $k_{\text{wall}}$. It reports, in force units, how far a flagellum tip has penetrated the contact buffer, and its only purpose is to set the tension pinned at the input edge. Its scale follows from the modeling choices $k_{\text{wall}}$ and $d_{\text{buf}}$;
no hydrodynamic calculation enters it. The resulting ceiling
$k_{\text{wall}}d_{\text{buf}}=0.3\,$pN does coincide with the propulsive force of a
single filament ($0.2$--$0.3\,$pN, S4.5), aligning with the bound on how hard a flagellum can press on a surface in Stokes flow. We define the contact duration $\tau_{\text{con}}$ as the time a flagellum spends above the contact gate; its median over near-wall contacts is $0.11\,$s.

\subsection*{S5.2 Trapping and escape}
The trap-and-release rule that defines the escape statistic is adapted from the
surface-alignment model of Junot \textit{et al.}~\cite{Junot2022}. On first
reaching the near-wall region---taken to lie within $d_{\text{trap}} = 8\,\mu$m
of the wall, the near-surface distance used in~\cite{Junot2022}---the cell is
held with its heading aligned parallel to the wall. Following the same model, it
remains trapped through any subsequent tumble whose reorientation still points
into the wall, and is released only when a tumble produces a heading directed
sufficiently away from the wall.

The quantitative release threshold follows the escape-angle argument
of~\cite{Junot2022}: a released cell must cross the near-wall layer of thickness
$d_{\text{trap}}$ in the wall-normal direction before its next tumble---that is,
within one mean run time $\tau_{\text{run}}$---or it tumbles again and is
re-trapped. Let $\theta_{\rm f}$ be the angle between the post-tumble heading and the
wall, so that the speed directed away from the wall is $v_{\text{run}}\sin\theta_{\rm f}$.
Escape within a single run then requires
\begin{equation}
  v_{\text{run}}\,\tau_{\text{run}}\,\sin\theta_{\rm f} \;\ge\; d_{\text{trap}}
  \Longrightarrow
  \sin\theta_{\rm f} \;\ge\; \sin\theta_c \;\equiv\; \frac{d_{\text{trap}}}{v_{\text{run}}\,\tau_{\text{run}}},
  \label{seq:escape}
\end{equation}
which is the criterion applied at each near-wall tumble: the cell is released
when its heading points away from the wall and $|\sin\theta_{\rm f}| > \sin\theta_c$. We take $\tau_{\text{run}}$ to be the mean
run duration produced by the switching calibration at baseline activity
(\cref{fig:calibration}), $\langle\tau_{\text{run}}\rangle \approx 1.0\,$s
($\langle f_{\text{tum}}\rangle = 0.87\,$Hz and
$\langle\tau_{\text{tum}}\rangle = 0.12\,$s give
$\langle\tau_{\text{run}}\rangle \approx 1/\langle f_{\text{tum}}\rangle
- \langle\tau_{\text{tum}}\rangle \approx 1.0\,$s). With
$d_{\text{trap}} = 8\,\mu$m and $v_{\text{run}} = 20\,\mu$m/s, this yields
$\sin\theta_c \approx 0.39$ and a mean run length
$v_{\text{run}}\langle\tau_{\text{run}}\rangle \approx 20\,\mu$m; the
simulation uses $|\sin\theta_{\rm f}| > 0.4$ (rounded).

\begin{figure}[tbp]
   \centering
    \includegraphics[width=\linewidth]{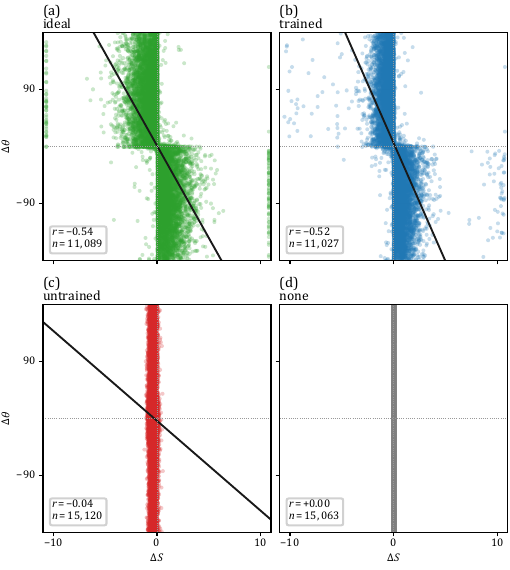}
\caption{\textbf{Sign-based steering coupling across four wall-response variants.}
Each panel pools signed stator asymmetry $\Delta S=S_L-S_R$ against signed
post-tumble angle $\Delta\theta$ for all near-wall tumbles in one variant (ideal
controller, trained, untrained, no network; $n$ annotated). Black line: linear
fit; $r$, pooled Pearson correlation. Trained ($r=-0.51$) and ideal ($r=-0.54$)
show anticorrelation: leftward $\Delta S$ drives rightward turns. Untrained
($r\approx-0.13$) and no-network ($r\approx0$) lack usable directional information.
This is the pooled counterpart of the trial-averaged bars in \cref{fig:pop-escape}(a)
and the trained/untrained overlay in \cref{fig:pop-escape}(e).}
\label{fig:steer-coupling-si}
\end{figure}

\subsection*{S5.3 Escape-statistics variants}
\begin{figure}[H]
   \centering
    \includegraphics[width=\linewidth]{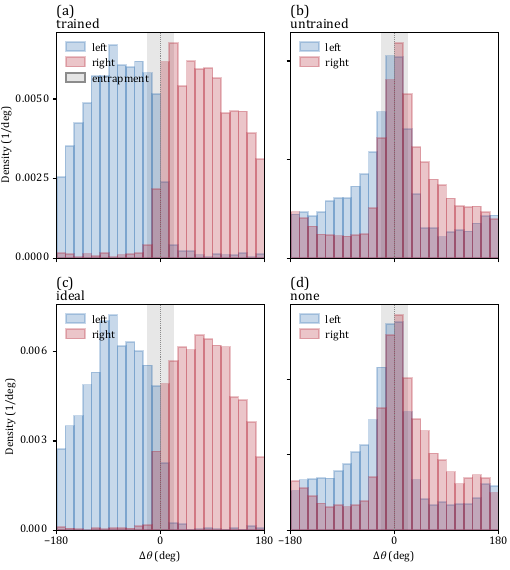}
\caption{\textbf{Impact-side resolution of the post-tumble angle, by variant.}
Signed tumble angle $\Delta\theta$ for near-wall tumbles, split by contact side
(blue: left impact; red: right impact), for (a)~trained network, (b)~untrained,
(c)~ideal controller, (d)~no network. Gray band: parallel-entrapment zone
$|\Delta\theta|<24^\circ$, where a tumble fails to redirect the cell out of the
near-wall layer. When routing is active, left- and right-impact distributions
separate to opposite signs of $\Delta\theta$, turning away from the struck wall;
without routing the distributions overlap. The escape-aligned quantity
$\Delta\theta_{\mathrm{esc}}$ in \cref{fig:pop-escape}(b) folds these raw
distributions onto a common away-from-wall axis by multiplying $\Delta\theta$ by
the sign of the required escape turn.}
\label{fig:panel-d-delta}
\end{figure}

\Cref{fig:pop-escape} compares four allostery pathways on an otherwise
identical wall experiment: same channel geometry, trapping rule, agent
dynamics, and contact threshold ($0.01\,$pN). Across the four variants, only the route from wall load to
$\sigma_{\text{PG},i}$ (and hence to $k_{\text{off},i}$ via \cref{seq:koff})
differs.

\begin{itemize}[nosep]
  \item \textbf{Trained network.}
  The quasistatic PG network reconstructed from the canonical trained
  checkpoint supplies $\sigma_{\text{PG}}$. On wall contact, contacting motors'
  input edges are pinned with tensions built from $F_{\text{wall}}$; the network is equilibrated and readout-edge tensions are passed to the off-rate model of \cref{seq:koff}.
  No substitute transfer rule is active.

  \item \textbf{Untrained network.}
  The same topology, geometry, and motor--edge assignment as the trained case,
  but bond stiffnesses and rest lengths are reset to their pre-training values. Wall forces still drive a
  quasistatic solve; only the learned bond field is absent.

  \item \textbf{Ideal controller.}
  The PG solve is bypassed. Instead, an analytic rule maps the instantaneous
  wall-force pattern directly to motor readout tensions. Using the same
  left/right split as elsewhere ($y>0$ vs.\ $y\le0$), let
  $F_L=\sum_{i:y_i>0}F_{\text{wall},i}$ and
  $F_R=\sum_{i:y_i\le0}F_{\text{wall},i}$ (motors below the
  contact threshold are excluded). The sign
  $\mathrm{sign}(F_L-F_R)$ identifies the impacted
  hemisphere. The rule then assigns, to every motor on the impacted side, the
  PG tension that would yield full stator occupancy ($S_i\to N_{\text{stat}}$)
  under the Bell inversion, and to every motor on the unimpacted side, the
  tension that would yield $S_i\to0$. This impacted-side assignment requires driving $k_{\text{off},i}\to0$, which the geometric channel cannot do on its own: $\sigma_{\text{comp},i}\ge0$ (\cref{seq:koff}) means it can only raise $k_{\text{off},i}$, never lower it. The ideal controller is therefore an upper bound in two senses, not one: no elastic-propagation loss through the sacculus, and an idealized bidirectional actuator that stabilizes the impacted side as well as destabilizing the unimpacted one, a capability the compression-only actuator (Model) does not have.

\item \textbf{No network.}
 A null baseline: the PG network is omitted and the readout tensions are set identically to zero ($\sigma_{\text{PG},i}=0$), so $k_{\text{off},i}$ depends only on the baseline load stabilizing-bond channel and no wall-derived allosteric signal reaches the motors.
\end{itemize}

\subsection*{S5.4 Carry-over of stator depletion between contacts}
\begin{figure}[tbp]
   \centering
    \includegraphics[width=\linewidth]{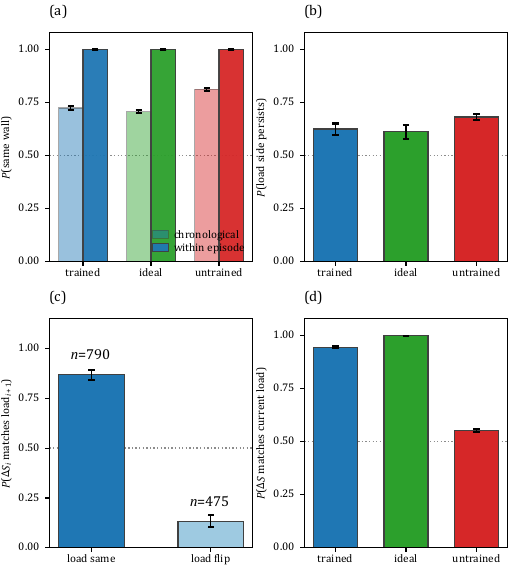}
\caption{\textbf{Carry-over of stator depletion between consecutive near-wall contacts.} All panels pool ordered pairs of consecutive near-wall tumbles from the runs of \cref{fig:pop-escape}; error bars, $95\%$ CI. \textbf{(a)}~Probability that two consecutive tumbles occur at the same wall, scored chronologically over a whole trial (light) and within a single entrapment episode (dark); within an episode the probability is $1$. \textbf{(b)}~Probability that the loaded hemisphere is the same for both tumbles of a within-episode pair. \textbf{(c)}~Trained network: probability that the earlier contact's $\Delta S$ has the polarity the later contact requires, split by whether the load side persisted or flipped. \textbf{(d)}~Probability that $\Delta S$ has the correct polarity for the contact on which it is measured, by variant. Panel~(a) shows that repeated same-wall contact is guaranteed within an episode; (b) and (c) show that this does not by itself keep the stale asymmetry correctly signed; (d) shows that the trained wall resolves the present load regardless.}
\label{fig:stale-carryover}
\end{figure}

Stator recovery is slow compared with the interval between near-wall contacts (S4.3), so depletion produced at one contact is still present at the next, and the pooled escape statistic of \cref{fig:pop-escape} includes those contacts. We score every ordered pair of consecutive near-wall tumbles in the runs of \cref{fig:pop-escape} on four quantities (\cref{fig:stale-carryover}): whether the two tumbles occur at the same wall, whether the loaded hemisphere is the same, whether the earlier $\Delta S$ has the polarity the later contact requires, and whether $\Delta S$ has the correct polarity for the contact on which it is measured. The load side is a cell-frame quantity, $\mathrm{sign}(F_L-F_R)$ with the left/right split of S5.3.
Consecutive tumbles within an entrapment episode always occur at the same wall (\cref{fig:stale-carryover}a), which follows from the channel geometry: the cell remains on one side until it escapes. Scored chronologically over a whole trial, including re-trapping after an escape, same-wall pairs remain common but not universal. Same wall does not imply the same loaded hemisphere, because a cell that reorients while remaining at one wall presents the opposite side; for the trained network the load side persists on $0.63$ of within-episode pairs (\cref{fig:stale-carryover}b). Comparing the earlier $\Delta S$ against the polarity the later contact requires, the stale asymmetry is correctly signed on $0.87$ of the pairs where the load side persists ($n=790$) and $0.13$ of the pairs where it flips ($n=475$), a weighted $0.59$ overall (\cref{fig:stale-carryover}c).
Routing on the present contact is a separate measurement and is unambiguous: the trained network assigns $\Delta S$ the polarity the current load requires on $0.95$ of near-wall tumbles and the ideal controller on essentially all of them, against $0.55$ for the untrained lattice (\cref{fig:stale-carryover}d). The non-decay of the pooled advantage is therefore consistent with repeated same-wall contact, and is accounted for by re-reading of the present load rather than by memory of the previous one.

\section*{S6. Mechanical routing under simultaneous loading}

The single-contact sweep of \cref{fig:steer-saturation}(a) applies a load to one side of the cell at a time. To test whether the trained mechanical pathways retain their directional response under simultaneous loading, we apply forces to both motor input edges in a single mechanical solve (S5.1). The left-side ($y>0$) input is held at $F_L=0.15\,\mathrm{pN}$, while the right-side ($y\leq0$) input is varied over $F_R\in[0,0.3]\,\mathrm{pN}$. The same protocol is applied to the untrained network. For the trained network, the difference in mean compression between the two sides,
$\Delta\sigma_{\text{comp}}=\langle\sigma_{\text{comp}}\rangle_L
-\langle\sigma_{\text{comp}}\rangle_R$, varies monotonically with $F_R$, from $-0.121\,\mathrm{pN}$ at $F_R=0$ to $-0.021\,\mathrm{pN}$ at equal loading, before becoming positive at larger $F_R$. The residual asymmetry at equal loading reflects
the asymmetric geometry and mechanical response of the disordered network. By contrast, the untrained network exhibits negligible
compression differences. The contact gate at $F_R=0.01\,\mathrm{pN}$ changes which motors contribute to the readout, rather than the underlying linear elastic response. The measured far-side routing ratio remains $0.905$ over the tested loading range, consistent with the single-contact result of \cref{fig:steer-saturation}(a).

\section*{S7. Parameters}

\begin{table*}[t]
\centering
\caption{Agent and wall parameters for the near-wall experiment. Justification: a
citation alone denotes a reported value; ``calc.'' a value derived from cited
quantities; ``calib.'' a constant tuned to reproduce the baseline \Ecoli{}
run--tumble statistics (\cref{fig:calibration}); ``choice'' a modeling choice
(``cf.'' = chosen within a cited range); ``const.'' a physical constant.}
\label{tab:bio-params}
\small
\begin{tabular}{@{}lllll@{}}
\toprule
\textbf{Symbol} & \textbf{Quantity} & \textbf{Value} & \textbf{Unit} & \textbf{Justification}\\
\midrule
\multicolumn{5}{@{}l}{\textit{Motor and stator mechanics}}\\
$N_{\text{mot}}$    & Flagellar motors per cell & 4    & ---            & \cite{Mears2014} \\
$N_{\text{stat}}$    & Max stator number         & 11   & ---            & \cite{Reid2006,Leake2006, Wang2022} \\
$r_{\text{mot}}$     & Rotor radius              & 22.5 & nm             & \cite{Francis1994} \\
$\omega_{\max}$        & Zero-load motor speed     & 1885 & rad/s          &~\cite{Wang2022} \\
$\zeta_r$              & Rotational drag           & 6.0  & pN$\cdot$nm$\cdot$s & \cite{Darnton2007}, calc. \\
\midrule
\multicolumn{5}{@{}l}{\textit{Stator off-rate (Bell form, \cref{seq:koff})}}\\

$\delta_{\text{LS}}$                 & Load-stabilizing bond distance  & 2.48 & nm              & \cite{Chawla2017} \\
$\delta_{\text{geo}}$ & Compression distance & 1.5  & nm              & choice \\
$\beta$                  & $1/k_BT$ (300\,K)    & 0.24 & (pN$\cdot$nm)$^{-1}$ & const. \\

$\tau_{\text{stat}}$ & Torque per stator (stall) & 250 & pN$\cdot$nm & \cite{Wang2022} \\

$k_{\text{on}}$ & Stator on-rate & 0.01 & s$^{-1}$ & choice; cf.\ \cite{Wadhwa2021} \\

$k_{\text{off}}^{\text{LS}}$ & Load stabilizing-bond baseline off-rate & 0.057 & s$^{-1}$ & \cite{Wadhwa2022} \\
$k_{\text{off}}^{\text{geo}}$   & Geometric baseline off-rate  & 600   & s$^{-1}$ & choice \\
\midrule
\multicolumn{5}{@{}l}{\textit{Run--tumble switching (\cref{seq:switching}; calibrated)}}\\
$h$                 & Switching Hill coefficient & 2.5         & ---      & calib. \\
$K_{1/2}$           & Switching half-saturation  & 1.25        & ---      & calib. \\
$k_{+},k_{-}$       & Switching prefactors       & 1.66,\,11.6 & s$^{-1}$ & calib. \\
$h_0$               & Baseline receptor activity & 1.3         & ---      & calib. \\
$N_{\text{th}}$ & CW motors for a tumble     & 2           & ---      & choice; cf.\ \cite{Mears2014} \\
$v_{\text{run}}$    & Run speed                  & 20          & $\mu$m/s & \cite{Berg1972,Turner2000} \\
$\Delta t$          & Time step                  & 0.01        & s        & choice \\
\midrule
\multicolumn{5}{@{}l}{\textit{Flagellar geometry and fluid (\cref{seq:omega}, S4.5)}}\\
$R_n,\,\beta_n$       & Normal helix radius, pitch & 0.175,\,25.5 & $\mu$m,\,deg & \cite{Dvoriashyna2021} \\
$R_{sc},\,\beta_{sc}$ & Semi-coiled radius, pitch  & 0.25,\,$-55$ & $\mu$m,\,deg & \cite{Dvoriashyna2021} \\
$L$                   & Filament contour length    & 10           & $\mu$m       & \cite{Turner2000} \\
$h_f$                 & Filament half-thickness    & 20           & nm           & \cite{Dvoriashyna2021} \\
$\mu$                 & Fluid viscosity            & $10^{-3}$    & Pa$\cdot$s   & \cite{Dvoriashyna2021} \\
\midrule
\multicolumn{5}{@{}l}{\textit{Wall and trapping (S5)}}\\
---               & Channel height               & 100  & $\mu$m       & choice \\
---               & Channel width                & 500  & $\mu$m       & choice \\
$k_{\text{wall}}$ & Wall contact stiffness       & 1.5  & pN/$\mu$m    & code parameter \\
$d_{\text{trap}}$ & Trap engage distance         & 8    & $\mu$m       & \cite{Junot2022} \\
$d_{\text{buf}}$  & Contact buffer width         & 0.2  & $\mu$m       & choice \\
$t_{\text{pin}}$  & Max.\ contact signal         & 0.3  & pN           & $k_{\text{wall}}d_{\text{buf}}$; also training input \\
$\sin\theta_c$    & Escape threshold             & 0.40 & ---          & \cref{seq:escape}, \cite{Junot2022} \\
---               & Min.\ wall force (PG step)   & 0.01 & pN           & choice \\
$c$               & Contact-to-anchor tension factor & 1.0 & ---  & choice \\
\bottomrule
\end{tabular}
\end{table*}

\begin{table*}[t]
\centering
\caption{Network and training parameters for the canonical run. The body is a
two-dimensional disordered (bidisperse, jammed) packing on a periodic torus with
Hookean bonds (S1); training uses the additive MmL learning rule
($\Delta k_b = -\gamma_l\,t^{\rm FF}_b t^{\rm FB}_b$) on a
linear-response solver (S2).}
\label{tab:net-train-params}
\small
\begin{tabular}{@{}llll@{}}
\toprule
\textbf{Quantity} & \textbf{Symbol} & \textbf{Value} & \textbf{Note}\\
\midrule
\multicolumn{4}{@{}l}{\textit{Network}}\\
Nodes              & $N$                 & 512                & \\
Box                & $L_x\times L_y$     & $10\times10$       & lattice units \\
Packing fraction   & $\phi$              & 0.84               & \\
Bidispersity ratio & ---                 & 1.4                & \\
Connection cutoff  & ---                 & 1.3                & in $r_{ij}/\sigma_{ij}$ \\
Bonds              & $N_b$               & 1497               & \\
Mean coordination  & $z=2N_b/N$          & 5.85               & $>z_c=4$ \\
Initial stiffness  & $k_0$               & $10^4$             & uniform \\
Stiffness bounds   & $k_{\min},k_{\max}$ & $10^{-10},10^{10}$ & effectively unbounded \\
\midrule
\multicolumn{4}{@{}l}{\textit{Training (MmL)}}\\
Learning rate      & $\gamma_l$ & $1.11\times10^{8}$ & $\Gamma=\gamma_l t_{\text{pin}}^{2}/k_0=10^{3}$ \\
Training input     & $t_{\text{pin}}$ & 0.30 & pN; $=k_{\text{wall}}d_{\text{buf}}$ (S5.1) \\
Training target    & $t^{D}$  & $-0.12$ & pN; ratio $-0.4$ \\

Tasks (cyclic)     & ---      & 8         & one per cycle \\
Cycles             & ---      & 3000      & \\
Convergence tol.   & ---      & $10^{-3}$ & on $\eta$ \\
Seed               & ---      & 42        & \\
\bottomrule
\end{tabular}
\end{table*}
\end{document}